\documentclass[11pt,tighten,nofootinbib,amssymb,floatfix]{article}
\usepackage[footskip=0.15in, margin=.75in]{geometry}

\usepackage{t1enc}
\usepackage{mathrsfs}
\usepackage{graphicx}
\usepackage{multirow}
\usepackage{hyperref}
\usepackage{bbm}
\usepackage{amsmath}
\usepackage{arydshln}
\usepackage{tikz}
\usepackage{authblk}

\usepackage{mathtools}

\newcommand{\be}{\begin{equation}}
\newcommand{\ee}{\end{equation}}
\newcommand{\bea}{\begin{eqnarray}}
\newcommand{\eea}{\end{eqnarray}}

\newcommand{\MSbar}{{\overline{\rm MS}}}

\def\smn{{\sigma_{\mu\nu}}}

\begin{document}

\enlargethispage{1\baselineskip}

\title{Perturbative Renormalization of quasi-PDFs}
\author{Martha Constantinou $^{a}$, Haralambos Panagopoulos $^b$
\footnote{Electronic address: marthac@temple.edu, haris@ucy.ac.cy}}

\vskip 0.25cm
\affil{
$^a$ {\small\it{Department of Physics,  Temple University,  Philadelphia,  PA 19122 - 1801,  USA}}\\
$^b$ {\small\it{Department of  Physics,  University  of Cyprus,  POB  20537,  1678  Nicosia,  Cyprus}}}

\date{}

\maketitle

\abstract{
In this paper we present results for the renormalization of gauge invariant 
nonlocal fermion operators 
which contain a Wilson line, to one-loop level in lattice perturbation theory. Our calculations have
been performed for Wilson/clover fermions and a wide class of Symanzik improved gluon actions.

The extended nature of such `long-link' operators results in a nontrivial renormalization, 
including contributions which diverge linearly as well as logarithmically 
with the lattice spacing, along with additional finite factors. 

On the lattice there is also mixing among certain subsets of these nonlocal operators; we 
calculate the corresponding finite mixing coefficients, which are necessary in order to
disentangle individual matrix elements for each operator from lattice simulation data. 
Finally, extending our perturbative setup, we
present non-perturbative prescriptions to extract the linearly divergent contributions.}

\section{Introduction}

Parton distribution functions (PDFs) provide important information on
the quark and gluon structure of hadrons; at leading twist, they
give the probability of finding a specific parton in the hadron
carrying certain momentum and spin, in the infinite momentum frame.
Due to the fact that PDFs are light-cone correlation functions, they
cannot be computed directly on a Euclidean lattice. Nevertheless, there is an alternative
approach, proposed by X. Ji \cite{Ji:2013dva}, involving the computation of quasi-distribution functions, which are
accessible in Lattice QCD. This formalism provides a promising means
of studying quark distribution functions in nucleons, given that, for large
momenta, one can establish connection with the physical PDFs through
a matching procedure. Exploratory studies of the quasi-PDFs reveal 
promising results for the non-singlet operators for the unpolarized, helicity
and transversity cases~\cite{Lin:2014zya,Alexandrou:2015rja,Chen:2016utp,Alexandrou:2016jqi}.

A standard way of extracting quasi-distribution functions in lattice simulations involves computing hadronic matrix elements of certain 
gauge-invariant nonlocal operators; the latter are made up of a product of an anti-quark field at position $x$, 
possibly some Dirac gamma matrices, a path-ordered exponential of the gauge field (Wilson line) along a path
joining points $x$ and $y$, and a quark field at position $y$ (particular cases are defined in Eq.~(\ref{Oper}) of the following Section). 
Given the extended nature of such operators, an endless variety of them, with 
different quantum numbers, can be defined and studied in lattice simulations and in phenomenological models. In pure gauge theories, 
prototype nonlocal operators are path-ordered exponentials along closed contours (Wilson loops); the contours may be smooth, but they may 
also contain angular points (cusps) and self-intersections. 

The history of investigations of nonlocal operators in gauge theories goes back a long time, including seminal
work of Mandelstam~\cite{PhysRev.175.1580} and Polyakov~\cite{POLYAKOV1979247}, and encompassing several 
complementary viewpoints~\cite{MAKEENKO1979135,Witten1989629}. In particular, the renormalization of Wilson loops
was studied perturbatively, in dimensional regularization (DR) using a $D$-dimensional spacetime, for smooth
contours~\cite{Dotsenko:1979wR} as well as for contours containing singular points~\cite{Brandt:1981kf}. 
Using arguments valid to all orders in perturbation theory, it was shown that smooth Wilson loops
in DR are finite functions of the renormalized coupling, while the presence of cusps and
self-intersections introduces logarithmically divergent multiplicative renormalization factors; at
the same time, it was shown that other regularization schemes are expected to lead to further
renormalization factors $Z$ which are linearly divergent with the dimensionful ultraviolet cutoff
$a$:
\be
Z = e^{\displaystyle - c\,L/a}\,,
\ee
where $c$ is a dimensionless quantity and $L$ is the loop length.

Certain nonlocal operators have also been studied extensively via lattice simulations in the past; typical
examples are products of open plaquettes at points $x$ and $y$, with varying orientations, joined
with two Wilson lines running in both directions between $x$ and
$y$~\cite{DiGiacomo:1992hhp,DiGiacomo:1996bbx}. Lattice results on matrix
elements of such operators have found extensive use in the description of 
chromoelectric and chromomagnetic field correlations, in phenomenological models
of the strong interactions (see, e.g.,~\cite{DiGiacomo:2000irz}, \cite{Simonov:2007dn}, \cite{Giordano:2009vs}, etc.); however, the renormalization
properties of these operators on the lattice need to be further explored.

At present, several aspects of Ji's approach are being investigated: The matching between quasi-PDFs and
physical PDFs~\cite{Xiong:2013bka,Ma:2014jla,Ma:2014jga,Chen:2016fxx,Li:2016amo},
the relation with transverse momentum-dependent parton distributions 
(TMDs)~\cite{Musch:2011er,Engelhardt:2015xja,Radyushkin:2016hsy,Radyushkin:2017ffo} and
the extraction of the linear divergence through studies of the static quark potential
\cite{Musch:2010ka,Ishikawa:2016znu,Chen:2016fxx}. Furthermore, there has been recent progress
towards the investigation of the logarithmic divergences \cite{Ishikawa:2016znu,Monahan:2016bvm,Carlson:2017gpk},
the demonstration that the quasi-PDF extracted from a Euclidean correlation function is the same matrix
element as that determined from the Lehmann-Symanzik-Zimmermann (LSZ) reduction formula in Minkowski 
spacetime~\cite{Briceno:2017cpo}, 
as well as the quark-in-quark quasi-PDF in lattice perturbation theory~\cite{Xiong:2017jtn}. 
There are several obstacles which need to be overcome before a transparent picture of PDFs 
can emerge via this approach; one such obstacle is clearly the intricate renormalization behavior, 
which is the object of our present study.

The paper is organized as follows: In Section 2 we formulate the problem, providing the
definitions for the lattice action and for the operators which we set out to renormalize, along with
the renormalization prescription. Section 3 contains our calculations, performed both in dimensional
regularization and on the lattice; we address in detail new features appearing on the
lattice, such as contributions which diverge linearly and logarithmically with the lattice spacing,
and finite mixing effects allowed by hypercubic symmetry. We also provide a 
prescription for estimating the linear divergence using non-perturbative data and following arguments
from one-loop perturbation theory. In Section 4 we summarize our results, and point out some
open questions for future investigations.

\section{Formulation}

\subsection{Lattice Actions}

In the calculation we make use of the clover (Sheikholeslami-Wohlert)
fermion action~\cite{Sheikholeslami:1985ij}; we allow the clover parameter, 
$c_{\rm SW}$, to be free throughout the calculation, in order to ensure wider applicability of our results. Using standard notation, this 
action reads:
\bea
S_F=\hspace{-0.2cm} &-&\hspace{-0.2cm} {a^3\over 2}\sum_{x,\,f,\,\mu}\bigg{[}\bar{
 \psi}_{f}(x) \left( r - \gamma_\mu\right) U_{x,\, x+a\,\mu}\psi_f(x+a\,\mu) 
+\bar{\psi}_f(x+a\,\mu)\left( r + \gamma_\mu\right)U_{x+a\,\mu,\,x}\psi_{f}(x)\bigg{]}\nonumber \\
&+&\hspace{-0.2cm} a^4 \sum_{x,\,f} (\frac{4r}{a}+m^f_0)\bar{\psi}_{f}(x)\psi_f(x) 
-{a^5\over 4}\,\sum_{x,\,f,\,\mu,\,\nu} c_{\rm SW}\,\bar{\psi}_{f}(x)
\smn F_{\mu\nu}(x) \psi_f(x)\,,
\label{clover}
\eea
where $r$ is the Wilson parameter (henceforth set to 1), $f$ is a flavor
index, $\smn =[\gamma_\mu,\,\gamma_\nu]/2$ and $F_{\mu\nu}$ is the standard
clover discretization of the gluon field tensor, defined through:
\be
{\hat F}_{\mu\nu} \equiv {1\over{8}}\,(Q_{\mu\nu} - Q_{\nu\mu})\,,
\ee
where $Q_{\mu\nu}$ is given by the sum of the plaquette loops:
\begin{eqnarray}
Q_{\mu\nu} \hspace{-0.2cm}&=& \hspace{-0.2cm} U_{x,\, x+\mu}U_{x+\mu,\, x+\mu+\nu}U_{x+\mu+\nu,\, x+\nu}U_{x+\nu,\, x}
+ U_{ x,\, x+ \nu}U_{ x+ \nu,\, x+ \nu- \mu}U_{ x+ \nu- \mu,\, x- \mu}U_{ x- \mu,\, x} \nonumber \\
&+&\hspace{-0.2cm} U_{ x,\, x- \mu}U_{ x- \mu,\, x- \mu- \nu}U_{ x- \mu- \nu,\, x- \nu}U_{ x- \nu,\, x}
+ U_{ x,\, x- \nu}U_{ x- \nu,\, x- \nu+ \mu}U_{ x- \nu+ \mu,\, x+ \mu}U_{ x+ \mu,\, x}\,.
\end{eqnarray}
We are interested in mass-independent renormalization schemes, and
therefore we set the Lagrangian masses for each flavor, $m^f_0$\,, to their critical value;
for a one-loop calculation this corresponds to $m^f_0{=}0$. Such a choice simplifies the algebraic expressions but requires 
special treatment of potential IR singularities.

\vspace{0.75cm}
In the gluon sector we employ a 3-parameter family of Symanzik improved actions involving
Wilson loops with 4 and 6 links, defined as~\cite{Horsley:2004mx}:
\bea
\hspace{-1cm}
S_G=\frac{2}{g_0^2} \Bigl[ \hspace{-0.2cm}&c_0&\hspace{-0.2cm} \sum_{\rm plaq.} {\rm Re\,Tr\,}\{1-U_{\rm plaq.}\}
\,+\, c_1 \sum_{\rm rect.} {\rm Re \, Tr\,}\{1- U_{\rm rect.}\} 
\nonumber \\ 
+ \hspace{-0.2cm}&c_2&\hspace{-0.2cm} \sum_{\rm chair} {\rm Re\, Tr\,}\{1-U_{\rm chair}\} 
\,+\, c_3 \sum_{\rm paral.} {\rm Re \,Tr\,}\{1-U_{\rm paral.}\}\Bigr].
\label{Symanzik}
\eea
The only restriction for the coefficients $c_i$ is a normalization condition, which ensures the correct
classical continuum limit of the action:
\be
c_0 + 8 c_1 + 16 c_2 + 8 c_3 = 1.
\label{norm}
\ee
In this work we employ several values for the coefficients $c_i$, but for simplicity, we present numerical results 
for three choices widely used in numerical simulations. These are the
Plaquette (Wilson), tree-level Symanzik-improved and Iwasaki 
actions; the corresponding values of the coefficients are shown in
Table~\ref{tab1}. Note that the one-loop Feynman diagrams which appear
in our calculation do not 
involve pure gluon vertices and, thus, no vertex depends on $c_i$\,; furthermore, the gluon propagator depends only on three combinations 
of the Symanzik coefficients: 
\bea
&&C_0 \equiv c_0 + 8 c_1 + 16 c_2 + 8 c_3 \,=1,\nonumber \\
&&C_1 \equiv c_2 + c_3, \\
&&C_2 \equiv c_1 - c_2 - c_3\,.\nonumber
\eea
Therefore, with no loss of generality we set $c_2=0$.

\begin{table}
\begin{center}
\begin{minipage}{6cm}
\begin{tabular}{lr@{}lr@{}lr@{}l}
\hline
\hline
\multicolumn{1}{c}{Action}&
\multicolumn{2}{c}{$c_{0_{\Large{\phantom{A}}}}^{{\Large{\phantom{A}}}}$} &
\multicolumn{2}{c}{$c_1$} &
\multicolumn{2}{c}{$c_3$} \\
\hline
\hline\\[-1ex]
$\,\,$Plaquette               &  1&           &  0&         &  0&                  \\ [0.5ex]
$\,\,$Symanzik                &  5&/3         & -1&/12      &  0&              \\ [0.5ex]
$\,\,$Iwasaki                 &  3&.648       & -0&.331     &  0&              \\ [0.75ex]
\hline
\hline
\end{tabular}
\end{minipage}
\end{center}
\vspace{-0.3cm}
\caption{Input parameters $c_0$, $c_1$, $c_3$ for selected gluon actions presented in this paper.}
\label{tab1}
\end{table}

\subsection{Definition of Operators}

To establish notation, let us first write the operators we study in this work, which have the general form:
\be
\mathcal{O}_\Gamma\equiv \overline\psi(x)\,\Gamma\,\mathcal{P}\, 
e^{i\,g\,\int_{0}^{z} A_\mu(x+\zeta\hat\mu) d\zeta}\, \psi(x+z\hat{\mu})\,,
\label{Oper}
\ee
with a Wilson line of length $z$ inserted between the fermion fields in order to ensure gauge invariance.
In the limit $z\to 0$, Eq.~(\ref{Oper}) reduces to the standard ultra-local fermion bilinear operators. However,
the calculation of the Green's functions for $\mathcal{O}^\mu_\Gamma$
is for strictly $z\neq 0$\,: the appearance of contact terms beyond
tree level renders the limit $z\to 0$ nonanalytic.

We consider only cases where the Wilson line is a straight line along any one of the four perpendicular directions,
which will be called $\mu$. Without loss of generality we choose~\footnote{Thus $\mu = 1$ should be identified
  with the $z$ direction, which is conventionally chosen for the Wilson line in numerical studies.} $\mu = 1$. 
We perform our calculation for all independent combinations of Dirac matrices, $\Gamma$, that is:
\vspace*{.25cm}
\begin{equation}
\Gamma = \hat{1},\quad \gamma^5,\quad \gamma^\nu,\quad \gamma^5\,\gamma^\nu,\quad  \gamma^5\,\sigma^{\nu\rho},\quad \sigma^{\nu\rho}.
\end{equation}
In the above, $\rho\ne\mu$ and we distinguish between the cases in which the index $\nu$ is in the same direction as the Wilson line ($\nu=\mu$),
or perpendicular to the Wilson line ($\nu\neq\mu$). For convenience, the 16 possible choices of $\Gamma$ are separated
into 8 subgroups, defined as follows:
\be
\label{S}
\begin{array}{lclrcl}
  {S} &\equiv& \mathcal{O}_{\hat{1}}            &&&      \\[0.5ex]
  {P} &\equiv& \mathcal{O}_{\gamma^5}           &&&      \\[0.5ex]
  {V}_1 &\equiv& \mathcal{O}_{\gamma^1}         &&&       \\[0.5ex]
  {V}_\nu &\equiv& \mathcal{O}_{\gamma^\nu} \,,\qquad &\nu& = &2, 3, 4             \\[0.5ex]
  {A}_1 &\equiv& \mathcal{O}_{\gamma^5\gamma^1}   &&&       \\[0.5ex]
  {A}_\nu &\equiv& \mathcal{O}_{\gamma^5\gamma^\nu} \,,\quad\,\, &\nu& =& 2, 3, 4    \\[0.5ex]
  {T}_{1\nu} &\equiv& \mathcal{O}_{\sigma^{1\nu}}  \,,\quad\,\,\, &\nu& = &2, 3, 4   \\[0.5ex]
  {T}_{\nu\rho} &\equiv& \mathcal{O}_{\sigma^{\nu\rho}} \,,\quad\,\,\, &\nu, \rho& = &2, 3, 4    \,.
\end{array}
\ee
One might also consider an alternative definition for the ``tensor'' operators: $T'_{\kappa\lambda} \equiv
\mathcal{O}_{\gamma^5\sigma^{\kappa\lambda}}$\,. Such a definition is clearly redundant if one employs a
4-dimensional regularization, such as the lattice, since the $T'$ operators are just a renaming of the $T$ operators,
and they will thus renormalize identically; as it turns out, even in dimensional regularization, where different
treatments of $\gamma_5$ amount to different renormalization prescriptions (see Subsection~\ref{ConversionFactors}), the
renormalization of the above two sets of tensor operators remains identical.

\subsection{Renormalization Prescription}

We perform the calculation in both the dimensional (DR) and lattice (LR) regularizations, which allows one to extract the renormalization
functions directly in the continuum $\MSbar$-scheme. The setup of this process is extensively described in 
Refs.~\cite{Constantinou:2015ela,Alexandrou:2016ekb} and is briefly outlined below. 
As is common practice, we will consider mass-independent
renormalization schemes, so that fermion renormalized masses will be vanishing; for the one-loop lattice calculations
this implies that the Lagrangian masses must be set to zero.

In the LR calculation we encounter finite mixing for some pairs of operators (see Subsection~\ref{resultsLR}),  and, thus, 
here we provide the renormalization prescription in the presence of mixing between two structures, $\Gamma_1$ and $\Gamma_2$,
where one has, a $2\times2$ mixing matrix~\footnote{All renormalization functions, generically labeled $Z$, depend
on the regularization $X$ ($X$ = DR, LR, etc.) and on the renormalization scheme $Y$ ($Y$ = $\MSbar$, RI$'$, etc.) and
should thus properly be denoted as: $Z^{X,Y}$, unless this is clear from the context.} ($Z$). 
More precisely, we find mixing within each of the pairs: $\{S,\, V_1\}$, $\{A_2,\, T_{34}\}$, $\{A_3,\, T_{42}\}$, $\{A_4,\, T_{23}\}$, 
in the lattice regularization.
In these cases, the renormalization of the operators is then given by a set of 2 equations:
\begin{equation}
  \binom{{\cal O}_{\Gamma_1}^R}{{\cal O}_{\Gamma_2}^R} =
  \begin{pmatrix} Z_{11} & Z_{12} \\ Z_{21} & Z_{22} \end{pmatrix}^{-1}
  \binom{{\cal O}_{\Gamma_1}}{{\cal O}_{\Gamma_2}}\,.
\end{equation}
Once the mixing matrix $Z_{ij}$ is obtained through the perturbative calculation of certain Green's functions, as shown
below, it can be applied to non-perturbative bare Green's functions derived from lattice simulation data, in order to
deduce the renormalized, disentangled Green's functions for each of the two operators separately.
Such an application will be presented in a follow up publication using Twisted Mass fermions~\cite{Alexandrou:2017huk}.

The one-loop renormalized Green's function of operator
$\mathcal{O}_{\Gamma_i}$ can be obtained from the one-loop bare  
Green's function of $\mathcal{O}_{\Gamma_i}$ and the tree-level Green's function of  $\mathcal{O}_{\Gamma_j}$ ($j\ne
i$); this can be seen starting from the general expression:
\be
\langle \psi^R\,{\cal O}^R_{\Gamma_i}\,\bar \psi^R \rangle_{\rm amp} =
Z_\psi\,\sum_{j=1}^2 \, (Z^{-1})_{ij} \,\langle\psi\,{\mathcal O}_{\Gamma_j}\,\bar \psi \rangle _{\rm amp} \,,
\qquad \psi  = Z^{1/2}_\psi \psi^R\,,
\label{eq1}
\ee
where the renormalization matrix $Z$ and the fermion field renormalization $Z_\psi$ have the following perturbative
expansion: 
\be
Z_{ij} = \delta_{ij} + g^2 z_{ij} + \mathcal{O}(g^4)\,,\qquad 
Z_\psi = 1 + g^2 z_\psi + \mathcal{O}(g^4)\,.
\ee
Throughout this work, $g$ denotes the gauge coupling; the distinction between bare and renormalized coupling is
immaterial for the one-loop perturbative calculation. All Green's functions are intended to be amputated, and thus the
label ``${\rm amp}$'' will be dropped from this point on.

Once the $\MSbar$ renormalized Green's functions have been computed in DR (see Section~\ref{resultsDR}), the condition for extracting $Z^{LR,\,\MSbar}_{11}$ and $Z^{LR,\,\MSbar}_{12}$ is simply the requirement that renormalized Green's functions be
regularization independent:
\be
\langle \psi^R\,{\cal O}^R_{\Gamma_i}\,\bar \psi^R \rangle^{DR, \,\MSbar} = 
\left.\langle \psi^R\,{\cal O}^R_{\Gamma_i}\,\bar \psi^R \rangle^{LR, \,\MSbar}\right|_{a\to 0}\,.
\ee
Substituting the right-hand side of the above relation by the expression in Eq.~(\ref{eq1}), there follows:

 \be
\hspace*{-0.35cm}
\langle \psi^R\,{\cal O}^R_{\Gamma_1}\,\bar \psi^R \rangle^{DR, \,\MSbar}
-\langle \psi\,{\cal O}_{\Gamma_1}\,\bar \psi \rangle^{LR}=
g^2\, \Big(z_\psi^{LR,\,\MSbar}-z_{11}^{LR,\,\MSbar}\Big) \langle \psi \,{\cal O}_{\Gamma_1}\,\bar \psi \rangle^{\rm tree} 
- g^2\, z^{LR,\,\MSbar}_{12}  \langle\psi\,{\cal O}_{\Gamma_2}\,\bar \psi \rangle^{\rm tree} + \mathcal{O}(g^4).
\label{eq2}
\ee
The Green's functions on the left-hand side of Eq.~(\ref{eq2}) are the main results of this work, 
where $\langle \psi\,{\cal O}_{\Gamma_1}\,\bar \psi \rangle^{DR, \,\MSbar}$ is the renormalized Green's function for ${\cal O}_{\Gamma_1}$
which has been computed in dimensional regularization (Eqs.~(\ref{SP_R})-(\ref{T2Tp2_R})) and renormalized using the $\MSbar$-scheme, while
$\langle \psi\,{\cal O}_{\Gamma_1}\,\bar \psi \rangle^{LR}$ is the bare Green's function of ${\cal O}_{\Gamma_1}$ in LR. 
The difference of the aforementioned Green's functions is polynomial in the external momentum (of degree 0, in our case,
since no lower-dimensional operators mix); in fact, verification of this property constitutes a highly nontrivial check
of our calculations.
Thus, Eq.~(\ref{eq2}) is an appropriate definition of the momentum-independent renormalization functions, $Z_{11}$ and $Z_{12}$.
Note that in the absence of mixing ($Z_{12}=Z_{21}=0$), Eqs.~(\ref{eq1}), (\ref{eq2}) reduce to:
\be
\langle \psi^R\,{\cal O}^R_{\Gamma_1}\,\bar \psi^R \rangle^{DR, \,\MSbar} =
Z_\psi^{LR, \,\MSbar}\, \, (Z_{11}^{LR, \,\MSbar})^{-1} \langle\psi\,{\mathcal O}_{\Gamma_1}\,\bar \psi \rangle^{LR}\,,
\label{eq3}
\ee
\be
\langle \psi^R\,{\cal O}^R_{\Gamma_1}\,\bar \psi^R \rangle^{DR, \,\MSbar}
-\langle \psi\,{\cal O}_{\Gamma_1}\,\bar \psi \rangle^{LR}=
g^2\Big(z_\psi^{LR,\,\MSbar}-z_{11}^{LR,\,\MSbar}\Big) \langle \psi \,{\cal O}_{\Gamma_1}\,\bar \psi \rangle^{\rm tree}  + \mathcal{O}(g^4).
\label{eq4}
\ee

\vspace*{0.5cm}
Non-perturbative evaluations of the renormalization functions cannot be obtained directly in the $\MSbar$ scheme;
rather, one may calculate them in some appropriately defined variant of the RI$'$ (``modified
regularization-invariant'') scheme, and then introduce the corresponding conversion  
factors between RI$'$ and $\MSbar$. Here we propose a convenient RI$'$ scheme which can be 
applied non-perturbatively, similar to the case of the ultra-local fermion composite operators, with due attention to
mixing. Defining, for brevity: $\Lambda_{\Gamma_i} = \langle\psi\,{\mathcal O}_{\Gamma_i}\,\bar \psi \rangle$\,, and denoting the corresponding renormalized Green's functions by $\Lambda_{\Gamma_i}^{{{\rm RI}'}}$\,,
we require:
\be 
{\rm Tr}\Big[\Lambda_{\Gamma_i}^{{{\rm RI}'}} \, (\Lambda_{\Gamma_j}^{\rm tree})^\dagger\Big]_{q_\nu = {\bar q}_\nu} =  
{\rm Tr}\Big[\Lambda_{\Gamma_i}^{\rm tree} \, (\Lambda_{\Gamma_j}^{\rm tree})^\dagger\Big] \qquad \left(\ = 12
\delta_{ij}\ \right).
\label{riprime}
\ee
The factor of 12 above originates from the fact that the trace acts on both Dirac and color indices. The momentum of the
external fermion fields is denoted by $q_\nu$, and the four-vector
${\bar q}_\nu$ denotes the RI$'$ renormalization scale. We note that the magnitude of $\bar q$ alone is not
sufficient to specify completely the renormalization prescription: Different directions in $\bar q$ amount to different
renormalization schemes, which are related among themselves via finite renormalization factors. In what follows we will
select RI$'$ renormalization scale 4-vector to point along the direction $\mu=1$ of the Wilson line: $(\bar q, 0, 0, 0)$.
The tree-level functions
$\Lambda_{\Gamma_j}^{\rm tree}$ are given by:
\be
\Lambda_\Gamma^{\rm tree}= \Gamma\,e^{i\,q_\mu z}\,.
\ee

Using Eq.~(\ref{eq1}) we express Eq.~(\ref{riprime}) in terms of bare Green's functions, obtaining:
\be
\frac{1}{12}\, Z_\psi^{LR,{{\rm RI}'}}\,\sum_{k=1}^2 {(Z^{LR,{{\rm RI}'}})^{-1}}_{ik} \,  
{\rm Tr}\Big[\Lambda_{\Gamma_k} \, (\Lambda_{\Gamma_j}^{\rm tree})^\dagger\Big]_{q_\nu = \bar q_\nu} = \delta_{ij}\,, 
\label{eq6}
\ee
\be
Z_\psi^{LR,{{\rm RI}'}} = \frac{1}{12}\, {\rm Tr}\Big[S \, \left(S^{\rm tree}\right)^{-1}\Big]_{q^2 = {\bar q}^2} \,,
\ee
where $S$  is the bare quark propagator and $S^{\rm tree}$ is its tree-level value; the one-loop computation of $S$ can be
found, e.g., in Ref.~\cite{Constantinou:2009tr}. 

Eq.~(\ref{eq6}) amounts to four conditions for the four elements of the
matrix $Z^{LR,{{\rm RI}'}}$. As it was intended, it lends itself to a non-perturbative evaluation of $Z^{LR,{{\rm RI}'}}$,
using simulation data for $\Lambda_{\Gamma_k}$\,. 

Converting the non-perturbative, RI$'$-renormalized Green's functions $\Lambda_{\Gamma_i}^{{{\rm RI}'}}$ to the $\MSbar$
scheme relies necessarily on perturbation theory, given that the very definition of $\MSbar$ is perturbative in
nature. We write:
\bea
  \binom{{\cal O}_{\Gamma_1}^{{{\rm RI}'}}}{{\cal O}_{\Gamma_2}^{{{\rm RI}'}}} &=&
  (Z^{LR,{{\rm RI}'}})^{-1}\cdot
  \binom{{\cal O}_{\Gamma_1}}{{\cal O}_{\Gamma_2}}\,,\ \ 
  \binom{{\cal O}_{\Gamma_1}^\MSbar}{{\cal O}_{\Gamma_2}^\MSbar} =
  (Z^{LR,\MSbar})^{-1}\cdot
  \binom{{\cal O}_{\Gamma_1}}{{\cal O}_{\Gamma_2}} 
\Rightarrow \nonumber\\
  \binom{{\cal O}_{\Gamma_1}^\MSbar}{{\cal O}_{\Gamma_2}^\MSbar} &=&  (Z^{LR,\MSbar})^{-1} \cdot (Z^{LR,{{\rm RI}'}})\cdot
  \binom{{\cal O}_{\Gamma_1}^{{{\rm RI}'}}}{{\cal O}_{\Gamma_2}^{{{\rm RI}'}}} \equiv
 (\mathcal{C}^{\MSbar,{{\rm RI}'}})\cdot
  \binom{{\cal O}_{\Gamma_1}^{{{\rm RI}'}}}{{\cal O}_{\Gamma_2}^{{{\rm RI}'}}}.
\label{eq7}
\eea
The conversion factor $\mathcal{C}^{\MSbar,{{\rm RI}'}}$ is a $2{\times}2$ matrix in this case; it is constant
($q$-independent) and stays finite as the regulator is sent to its limit ($a\to 0$ for LR, $D\to 4$ for DR).
Most importantly, its value is independent of the regularization: 
\be
\mathcal{C}^{\MSbar,{{\rm RI}'}} = (Z^{LR,\MSbar})^{-1} \cdot (Z^{LR,{{\rm RI}'}}) = (Z^{DR,\MSbar})^{-1} \cdot (Z^{DR,{{\rm RI}'}}).
\label{convfactor}
\ee
Thus, the evaluation of $\mathcal{C}^{\MSbar,{{\rm RI}'}}$ can be performed in DR, where
evaluation beyond one loop is far easier than in LR; this, in a nutshell, is the advantage of using the RI$'$ scheme as an
intermediary. A further simplification originates from the fact that the DR mixing matrices $Z^{DR,\MSbar}$ and
$Z^{DR,{{\rm RI}'}}$ are both diagonal; as a result, $\mathcal{C}^{\MSbar,{{\rm RI}'}}$ turns out to be diagonal as well.  
We stress that in the case of ultra-local operators, the conversion factor depends
on the renormalized coupling and the ratio of the $\MSbar$ over the RI$'$ renormalization
scales ($\bar\mu/\bar q$); for the Wilson line operators on the other hand, the conversion
may (and, in general, will) depend on the length of the Wilson line and on the
individual components of the RI$'$ renormalization-scale four-vector, through the
dimensionless quantities $z{\bar q}_\nu$\,. 
Finally, let us also point out that, just as in the case of ultra-local operators, the non-perturbative evaluation of RI$'$
renormalization functions is performed at nonzero renormalized quark masses $m_q$\,; this would, in principle,
necessitate that the perturbative evaluation of $\mathcal{C}^{\MSbar,{{\rm RI}'}}$ also be performed at the same nonzero values
of $m_q$\,. However, given the near-critical values for light quark masses employed in present-day simulations,
$m_q/\bar\mu \ll 1$, and given the smooth dependence of $\mathcal{C}^{\MSbar,{{\rm RI}'}}$ on
$m_q$\,, only imperceptible changes are expected by setting $m_q\to 0$. 

Once the conversion factor is computed, the Green's functions $\Lambda_{\Gamma_i}^{{{\rm RI}'}}$ can be directly converted
to the $\MSbar$ 
scheme through:
\be
\binom{\Lambda_{\Gamma_1}^\MSbar}{\Lambda_{\Gamma_2}^\MSbar} = 
{Z_\psi^{LR,\MSbar}\over Z_\psi^{LR,{{\rm RI}'}}} \, (Z^{LR,\MSbar})^{-1} \cdot (Z^{LR,{{\rm RI}'}}) \cdot 
\binom{\Lambda_{\Gamma_1}^{{{\rm RI}'}}}{\Lambda_{\Gamma_2}^{{{\rm RI}'}}} =
{1\over \mathcal{C}_\psi^{\MSbar,{{\rm RI}'}}}\, 
 (\mathcal{C}^{\MSbar,{{\rm RI}'}})\cdot
  \binom{\Lambda_{\Gamma_1}^{{{\rm RI}'}}}{\Lambda_{\Gamma_2}^{{{\rm RI}'}}}.
\label{convGreen}
\ee
The fermion field conversion factor: $\mathcal{C}_\psi^{\MSbar,{{\rm RI}'}} \equiv Z_\psi^{LR,{{\rm RI}'}}/Z_\psi^{LR,\MSbar} 
= Z_\psi^{DR,{{\rm RI}'}}/Z_\psi^{DR,\MSbar}$ 
is a finite function of the renormalized coupling constant, and its value is
known well beyond one loop~\cite{Chetyrkin:1999pq}. 

In non-perturbative studies of Green's functions with physical nucleon states, performed via lattice simulations,
the external states are normalized in a way which does not involve the quark field renormalization $Z_\psi$\,; thus, the
only conversion factor necessary in this case is $\mathcal{C}^{\MSbar,{{\rm RI}'}}$.

\section{Calculation - Results}

\begin{figure}[h]
\centerline{\includegraphics[scale=1]{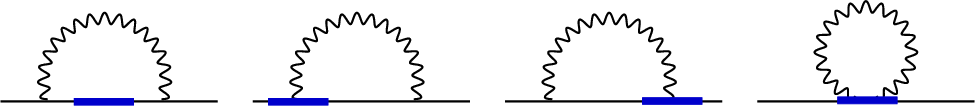}}
\centerline{\bf{d1\hspace*{3.5cm} d2\hspace*{4cm} d3\hspace*{3.5cm} d4}}
\vspace*{-0.3cm}
\begin{center}
\begin{minipage}{15cm}
\hspace*{3cm}
\caption{\small{Feynman diagrams contributing to the one-loop calculation of the Green's functions of operator $\mathcal{O}_\Gamma$.
The straight (wavy) lines represent fermions (gluons). The operator insertion is denoted by a filled rectangle.}}
\label{fig1}
\end{minipage}
\end{center}
\end{figure}

The Feynman diagrams that enter our one-loop calculations are shown in Fig.~\ref{fig1}, where 
the filled rectangle represents the insertion of any one of the nonlocal operators $\mathcal{O}_\Gamma$ with a Wilson
line of length $z$.  
Diagram d1 contains the 0-gluon vertex of the operator ($\mathcal{O}(g^0)$ of Eq.~(\ref{Oper})), whereas diagrams d2-d3
(d4) contain the corresponding 1-gluon (2-gluon) vertex. These diagrams will appear in our calculations in both
LR and DR, since all vertices are present in both regularizations, and since even the ``tadpole'' diagram d4 does not
vanish in DR, by virtue of the nonlocal nature of $\mathcal{O}_\Gamma$\,.
However, the LR calculation is much more challenging: The vertices of $\mathcal{O}_\Gamma$ are more complicated, and
extracting the singular parts of the Green's functions is a more lengthy and subtle procedure.

\subsection{Dimensional Regularization}
\label{resultsDR}
Let us first recall some of the essentials of dimensional regularization and of the $\MSbar$ scheme.
The computation is performed in $D$ Euclidean spacetime dimensions, where $D=4 -2\epsilon$ and 
$\epsilon$ is the regularizing parameter. Bare $n$-loop Green's functions in DR will be Laurent series
in $\epsilon$, of the form $\sum_{i = -n}^\infty c_i\,\epsilon^i$\,;
renormalization must eliminate all poles (negative powers) in $\epsilon$, before the limit $D\to 4$ can be taken.
The simplest scheme for the elimination of divergences is the modified minimal 
subtraction ($\MSbar$) scheme according to which the renormalization functions are defined to
only remove poles in $\epsilon$, without any finite parts~\footnote{Together
with this operation one must also
express every occurence of the dimensionful scale $\mu$ in terms of the $\MSbar$
renormalization scale $\bar\mu\equiv \mu \left(4\pi/e^{\gamma_E}\right)^{1/2}$,
where $\gamma_E$ is the Euler constant and $\mu$ appears in bare
Green's functions by virtue of the relation between the
$D$-dimensional bare coupling $g$ and the renormalized coupling $g^R$: $g = \mu^\epsilon\,Z_g\, g^R$.}. $\MSbar$ has become a reference scheme, and although
other schemes, such as RI$'$, are more suitable for non-perturbative calculations, an appropriate conversion factor is
applied to reach 
$\MSbar$. In this Section we present our results for the renormalization functions in the
$\MSbar$ scheme, and provide the conversion factor between RI$'$, as defined in Eq.~(\ref{riprime}),
and $\MSbar$.

\bigskip
For the extraction of UV divergences and determination of the poles in $\epsilon$, we follow the standard procedure of
introducing Feynman parameters which allow us
to perform the $D$-dimensional loop integrals~\cite{tHooft:1973wag}. However, unlike ordinary massless 2-point Green's
functions, in which the dependence on the external momentum $q$ is dictated purely on dimensional grounds, the results of
integration are now considerably more complicated; this is due to the appearance of both $q$ and the length $z$ in the 
integrands, along with a nontrivial dependence on the preferred direction
of the Wilson line. Just to give an example, we present one of the simpler integrals which
appear in the one-loop calculation:
\bea
{\int_0^z \hspace*{-0.1cm} d\zeta \hspace*{-0.1cm}} \int
{\hspace*{-0.1cm}\frac{dp^D}{(2\pi)^D}}\,\frac{e^{-i\,\zeta\,p_\mu}\,p_\mu}{{p}^{\,2}\,{(p+q)^2}}
\hspace*{-0.2cm}&=&\hspace*{-0.2cm}
  \frac{\Gamma(\frac{5{-}D}{2}) } {(4\pi)^{(D{-}1)/2}} \hspace*{-0.1cm}\int_0^z \hspace*{-0.1cm}d\zeta
\hspace*{-0.1cm}\int_0^1 \hspace*{-0.1cm} dx \hspace*{-0.1cm}\int_{-\infty}^\infty \hspace*{-0.1cm} dp_\mu \,
   e^{i \zeta (p_\mu {-} x q_\mu)} \, (p_\mu {-} x q_\mu) \,
(p_\mu^2 {+} q^2 x (1{-}x))^{\frac{D{-}5}{2}}\,
 \nonumber\\[1ex]
&&\hspace*{-2.6cm} = \frac{i}{16\pi^2} \Big[{1\over \epsilon} + 2 - \gamma_E
 + \log(4\pi/q^2) - 2 \int_0^1 dx \,e^{-i x z q_\mu} \, K_{0}(|z| \sqrt{ q^2x(1{-}x)} )\Big],
\label{IntDR}
\eea
where $p$ is the loop momentum, $q$ is the external momentum, $x$ is a Feynman
parameter to be integrated over, and $K_n$ is the modified Bessel
function of the second kind. In the first line of Eq.~(\ref{IntDR}),
integration was performed over the $D{-}1$ components of $p$
perpendicular to $p_\mu$\,; subsequently, a Laurent expansion was made,
keeping terms up to ${\cal O}(\epsilon^0)$.
We see that the $1/\epsilon$ pole part of this expression has a constant
coefficient; even though a potential $z$-dependence
would not come into conflict with renormalizability, it cannot arise,
since neither $q$ nor any renormalization scale can appear in the pole
part and, therefore, no dimensionless combination can be formed out of
$z$. The finite part, on the other hand, clearly exhibits a nontrivial
dependence on the dimensionless quantities: $zq$, $zq_\mu$, in
addition to
the standard logarithmic dependence on $q/\bar\mu$.

In a similar way we manipulate all integrals appearing in the computation in dimensional regularization. 
In Subsection~\ref{resultsLR} we will discuss the evaluation of the lattice counterpart of the above integral 
and describe the highly intricate process for the proper extraction of the UV divergences.

\subsubsection{Renormalization Functions}
\label{RenormalizationFunctionsDR}

The perturbative calculation has been performed in an arbitrary covariant gauge in order to see first-hand 
the gauge invariance of the renormalization functions; this serves as a consistency check of our calculation. 
The gauge fixing parameter, $\beta$, is defined such that $\beta{=}0 (1)$ corresponds to the Feynman (Landau) gauge. 
We find that $1/\epsilon$ terms arise from diagrams d2, d3 and d4, and thus d1 does not contribute
to the $\MSbar$ renormalization function of the operators under study. However, d1 contributes to the renormalized
Green's functions and to the conversion factors from RI$'$ to 
$\MSbar$. Below we present the ${\cal O}(1/\epsilon)$ contributions to 
$\Lambda_\Gamma = \langle\psi\,{\mathcal O}_\Gamma\,\bar \psi \rangle$ from each of the diagrams, including all
combinatorial factors:
\bea
\displaystyle
\Lambda^{d1}_\Gamma\Big{|}_{1/\epsilon} &=& 0 \\
\Lambda^{d2+d3}_\Gamma\Big{|}_{1/\epsilon} &=& 
\displaystyle\frac{g^2\, C_f}{16\,\pi^2}\,\frac{1}{\epsilon}\left(2- 2\beta\right) \Lambda_\Gamma^{\rm tree} \\
\Lambda^{d4}_\Gamma\Big{|}_{1/\epsilon} &=&  
\displaystyle\frac{g^2\, C_f}{16\,\pi^2}\,\frac{1}{\epsilon}\left(2 +\beta\right) \Lambda_\Gamma^{\rm tree}\,,
\eea
and thus:
\be
\langle \psi\,{\cal O}_{\Gamma}\,\bar \psi \rangle^{DR}\Big{|} _{1/\epsilon} = 
g^2\, \lambda_\Gamma \,\Lambda_\Gamma^{\rm tree}, \qquad
\lambda_\Gamma =\frac{C_f}{16\,\pi^2}\,\frac{1}{\epsilon}\left(4 - \beta \right)\qquad [C_f \equiv (N^2-1)/(2N)].
\label{eq5}
\ee
Note that to one-loop level in the DR calculation the pole parts are multiples of the tree-level values, which indicates
no mixing between operators of equal or lower dimension. Also, diagrams d2 and d3, the so-called `sail' diagrams, 
are symmetric and give the same contribution to the total Green's function. Another important characteristic of the 
${\cal O}(g^2)/\epsilon$ contributions is that they are operator independent, in terms of both the Dirac structure 
and the length of the Wilson line, $z$. 
Using Eq.~(\ref{eq5}) and $Z_\psi$ in DR~\cite{Gracey:2003yr}:
\be
Z^{DR,\,\MSbar}_\psi = 1  + g^2 z_\psi\,, \qquad z_\psi = \frac{C_f}{16\,\pi^2}\,\frac{1}{\epsilon}\,\Big(\beta - 1\Big)\,,
\ee
the $\MSbar$ condition to one loop reads:
\be
g^2\, z_\psi - g^2\, z_\Gamma + g^2\,\lambda_\Gamma = 0\,.
\ee
As expected from gauge invariance, we find a gauge independent renormalization function for the operators of Eq.~(\ref{Oper}):
\begin{equation}
Z^{DR,\,\MSbar}_\Gamma = 1 + \frac{3}{\epsilon}\, \frac{g^2\, C_f}{16\,\pi^2}\,,
\end{equation}
in agreement with Refs. ~\cite{Dorn:1986dt,Chetyrkin:2003vi}. Gauge invariance is not guaranteed in all schemes, 
since the Green's functions on which the renormalization condition is
imposed typically contain gauge variant renormalized external fields. Since the
$\MSbar$ scheme removes only the divergences, which are universal for all (gauge dependent and independent) Green's functions, 
it has to be gauge invariant.

While the independence of $Z^{DR,\,\MSbar}_\Gamma$ from the Dirac matrix insertion $\Gamma$ is a feature valid
to one-loop level, its independence from the length of the Wilson line $z$ is expected to hold to all orders in
perturbation theory; this, in essence, is due to the fact that the most dominant pole at every loop can depend neither on the
external momenta nor on the renormalization scale, thus there is no dimensionless $z$-dependent factor that could appear
in the pole part.

\subsubsection{Conversion Factors}
\label{ConversionFactors}

The Green's functions in DR are also very useful for the computation of the conversion factors between 
different renormalization schemes, and here we are interested in the RI$'$ scheme defined in Eq.~(\ref{eq6}). 
To one-loop level, the conversion factor is simplified to:
\be
\mathcal{C}_\Gamma^{\MSbar,{{\rm RI}'}} = 1 + g^2\, z_\Gamma^{DR, {{\rm RI}'}} - g^2 z_\Gamma^{DR, \MSbar}\,,
\ee
which we have computed for all operators shown in Eqs.~(\ref{S}). Note that our one-loop calculations do not depend on
the prescription which one adopts for extending $\gamma_5$ to $D$ dimensions (see, e.g.,
Refs.~\cite{Buras:1989xd,Patel:1992vu,Larin:1993tp,Larin:1993tq,Skouroupathis:2008mf,Constantinou:2013pba} 
for a discussion of four relevant prescriptions and some conversion factors among them); this is because (anti-)commutation
relations among $\gamma_5$ and $\gamma_\nu$ appear only in contributions which are finite as $\epsilon\to 0$. 
Also, the conversion factor is
the same for each of the following pairs of operators: Scalar and pseudoscalar, vector and axial ``parallel'' ($V_1,\ A_1$), vector and axial ``perpendicular''
($V_{2,3,4},\ A_{2,3,4}$), as well as for the tensor with and without $\gamma^5$\,;
furthermore, all components of $T_{\nu\rho}$ will have the same conversion factor, regardless of whether $\nu$ or
$\rho$ are parallel or perpendicular to the Wilson line. 
The general expression for $\mathcal{C}_\Gamma $ are shown in Eqs.~(\ref{C_SP})-(\ref{C_T}) for general gauge fixing parameter. 
They are expressed compactly in terms of the quantities $F_1(\bar q, z)$ - $ F_5(\bar q, z)$ and $G_1(\bar q, z)$ - $G_5(\bar q, z)$, 
which are integrals over modified Bessel functions of the second kind, $K_n$. These integrals are presented in Eqs.~(\ref{F1})-(\ref{G6}) 
of Appendix A. 
Just as in the example of Eq.~(\ref{IntDR}), the conversion factors depend on the dimensionless quantities $z\bar q$ and
$\bar q/\bar\mu$. The RI$'$ and $\MSbar$ renormalization scales ($\bar q$ and $\bar\mu$, respectively) have been left independent.

The Green's functions of operators with a Wilson line are complex, a property which is also valid for the non-perturbative matrix
elements between nucleon states (see, e.g. Ref.~\cite{Alexandrou:2016jqi}); this is also propagated to the conversion factor,
as can be seen in the following equations.

\bea 
\hspace*{-.65cm}
 \mathcal{C}_{S(P)} = 1 - \frac{g^2\,C_f}{16\,\pi^2} \hspace*{-0.4cm}&\Bigg(& \hspace*{-0.4cm}
    {-}7 {-} 4 \gamma_E {+} \log(16) -8 {F_2}+2 \bar q |z| (F_4-2 {F_5})  -3 \log
    \left(\frac{{\bar{\mu}^2}}{{\bar q^2}}\right)-(\beta +2) \log ({\bar q^2}
    {z^2})\nonumber \\
       &&\hspace*{-0.4cm} +\beta  \Big[ 2 {-} 2 \gamma_E {+} \log(4) -2 {F_1}-{\bar q^2} z^2\left(\frac{{F_1{-}F_2}}{2}-{F_3}\right)
    -2 \left({\bar q_\mu}^2+{\bar q^2}\right) G_3 +\bar q |z| F_4\Big]\nonumber \\
    &&\hspace*{-0.4cm}+ i \Bigg\{4 {\bar q_\mu}
    \Big(z(-{F_1{+}F_2}{+}{F_3}) {+}{G_1}\Big)
    +\beta  {\bar q_\mu} \Big[-2 (G_1 {-}G_2)
+\bar q \Big(2 ({G_4}{-}2{G_5})+ z |z|{F_5}\Big)\Big]\Bigg\}\Bigg)\quad
    \label{C_SP}
\eea

\bea
\hspace*{-.65cm}
 \mathcal{C}_{V_1 (A_1)} = 1 - \frac{g^2\,C_f}{16\,\pi^2} \hspace*{-0.4cm}&\Bigg(& \hspace*{-0.4cm}
 {-}7 {-} 4 \gamma_E {+} \log(16) {+}\frac{4 |z| \left(
    \left({\bar q^2}{+}{\bar q_\mu}^2\right){F_5}{-}{\bar q^2}{F_4}
    \right)}{\bar q} {+}4 {F_2} -3 \log
    \left(\frac{{\bar\mu^2}}{{\bar q^2}}\right){-}(\beta {+}2) \log ({\bar q^2}{z^2})\nonumber  \\
&&\hspace*{-0.4cm} {+}\beta  \Big[ 2 {-} 2 \gamma_E {+} \log(4) {-}\frac{2  {\bar q_\mu}^2
    |z|}{\bar q}{F_4}{-}2 {F_1}{+}z^2
    \left({\bar q_\mu}^2 ({F_3}{-}{F_1{+}F_2}){+}{\bar q^2}\frac{{F_1{-}F_2}
    }{2}\right){-}2 
    \left({\bar q^2}{+}{\bar q_\mu}^2\right){G_3}\Big]\nonumber  \\
&&\hspace*{-0.4cm}  {+}i \Bigg\{
4 {\bar q_\mu} (2 z ({F_1{-}F_2}{-}{F_3}) {+}{G_1})
{+}\beta  {\bar q_\mu} \Big[\bar q ( z |z|{F_5}{+}2
    ({G_4}{-}2{G_5})){-}2
    {G_1}{+}2 {G_2}\Big]\Bigg\} \Bigg)\qquad\qquad
    \eea

\bea
\hspace*{-.65cm}
\mathcal{C}_{V_\nu(A_\nu)} = 1 - \frac{g^2\,C_f}{16\,\pi^2} \hspace*{-0.4cm}&\Bigg(& \hspace*{-0.4cm}
 {-}7 {-} 4 \gamma_E {+} \log(16) +4 {F_2}+\frac{4  {\bar q_\nu}^2
    |z|}{\bar q}{F_5} -3 \log
    \left(\frac{{\bar\mu^2}}{{\bar q^2}}\right)-(\beta +2) \log ({\bar q^2}
    {z^2})\nonumber  \\
&&\hspace*{-0.4cm}  {+}\beta  \Big[ 2 {-} 2 \gamma_E {+} \log(4) {-}2
    \left(\frac{ {\bar q_\nu}^2 |z|}{\bar q}{F_4}{+}
    \left({\bar q^2}{+}{\bar q_\mu}^2\right){G_3}\right){-}2 {F_1}{+}z^2 \left({\bar q^2}
    \left({F_3}{-}\frac{{F_1{-}F_2}}{2}\right){+}{\bar q_\nu}^2
    ({F_1{-}F_2}{-}{F_3})\right)\Big]\nonumber  \\
&&\hspace*{-0.4cm}  {+}i \Bigg\{4 {\bar q_\mu}{G_1}+ \beta
    {\bar q_\mu} \Big[\bar q ( z |z|{F_5}+2
    ({G_4}-{2}{G_5}))-2
    {G_1}+2 {G_2}\Big]\Bigg\}\Bigg)
\eea

\bea
\hspace*{-.65cm}
\mathcal{C}_{T} = 1 - \frac{g^2\,C_f}{16\,\pi^2} \hspace*{-0.4cm}&\Bigg(& \hspace*{-0.4cm}
 {-}7 {-} 4 \gamma_E {+} \log(16)  {+}8 {F_2} {-}2 \bar q |z| ({F_4}{-}2 {F_5}) -3 \log
    \left(\frac{{\bar\mu^2}}{{\bar q^2}}\right){-}(\beta {+}2) \log ({\bar q^2}{z^2})\nonumber  \\
&&\hspace*{-0.4cm} 
    {+}\beta\Big[ 2 {-} 2 \gamma_E {+} \log(4) {-} \bar q |z|{F_4}{-}2 {F_1}{+}z^2 \left(({F_3}{-}{F_1{+}F_2})
    \left({\bar q_\mu}^2{+}{\bar q_\nu}^2\right){+}{\bar q^2}\frac{{F_1{-}F_2}
    }{2}\right){-}2 \left({\bar q^2}{+}{\bar q_\mu}^2\right){G_3}\Big] \nonumber  \\
&&\hspace*{-0.4cm} {+}i \Bigg\{
4 {\bar q_\mu} (z({F_1{-}F_2}{-}{F_3}) {+}{G_1})
{+}\beta {\bar q_\mu} \Big[\bar q ( z |z|{F_5}{+}2
    ({G_4}{-}2{G_5})){-}2
    {G_1}{+}2 {G_2}\Big]\Bigg\}\Bigg),
 \label{C_T}
\eea
where $q {\equiv} \sqrt{q^2} $. 

In Appendix B we present the $\MSbar$-renormalized Green's functions for each Wilson line operator. Starting from these
Green's functions, the conversion factors
Eqs.~(\ref{C_SP})-(\ref{C_T}) can be derived in a straightforward manner, but have been included for ease of reference.

Note that for a scale of the form $(\bar q,0,0,0)$ the one-loop Green's functions are a multiple of the 
tree-level value of the operator under consideration. The conversion factors take the form:
\bea 
\hspace*{-.65cm}
 \mathcal{C}_\Gamma = 1 + \frac{g^2\,C_f}{16\,\pi^2} \hspace*{-0.4cm}&\Bigg(& \hspace*{-0.4cm}
-3\,\log \left(\frac{{\bar{\mu}^2}}{{\bar q^2}}\right)+(\beta -1) + 
F_\Gamma(\bar{q} z) \Bigg)\,,
\label{C_Gamma}
\eea
where $F_\Gamma(\bar{q} z)$ are defined in Eqs.~(\ref{F_SP}) - (\ref{F_T}).

We remind the reader that the conversion factors
as defined in Eq.~(\ref{convfactor}) are to be multiplied by the $Z^{\MSbar}$ to give $Z^{{\rm RI}'}$. Alternatively,
one may obtain $Z^{\MSbar}$ by multiplying $Z^{{\rm RI}'}$ with $\mathcal{C}_\Gamma$ provided that $g^2 \to -g^2$, 
which is valid to one-loop level.

\begin{figure}[h]
\centerline{\includegraphics[scale=.285,angle=-90]{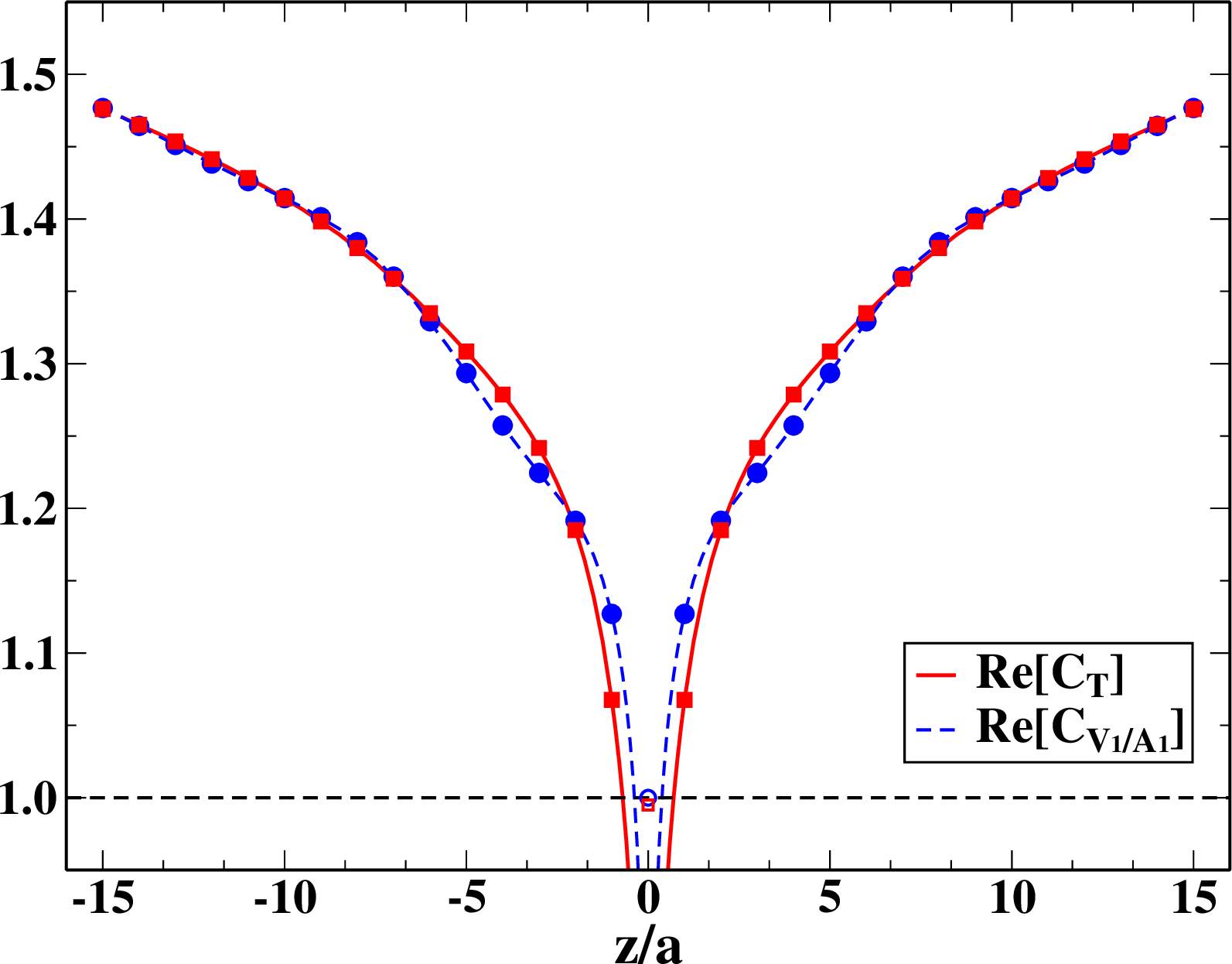}\quad
\includegraphics[scale=.285,angle=-90]{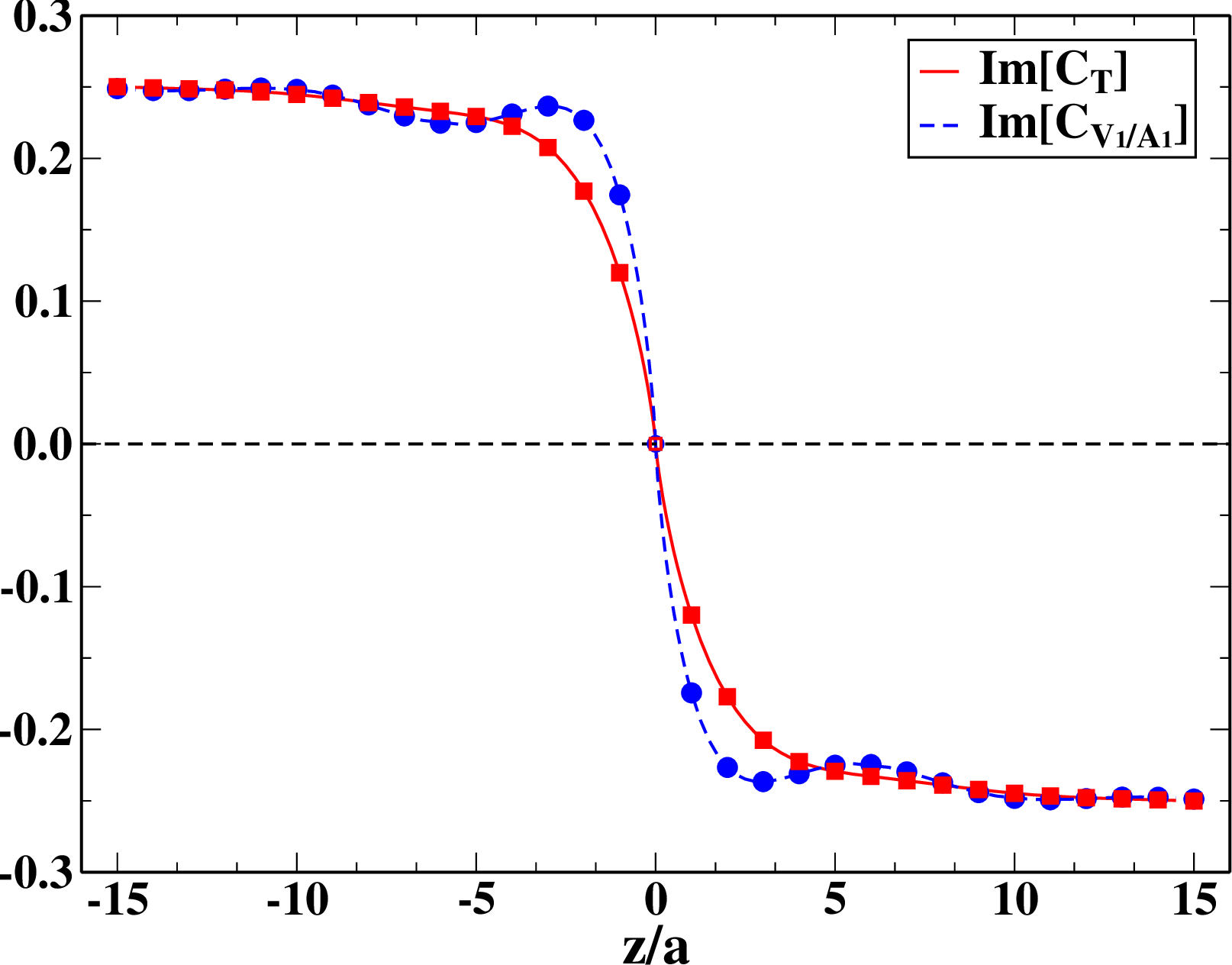}}
\vspace*{-0.3cm}
\begin{center}
\begin{minipage}{15cm}
\hspace*{3cm}
\caption{\small{Real (left panel) and imaginary (right panel) parts of the conversion factors for the operators $V_1$ and $T$ as a function of $z/a$ in the Landau gauge. 
The RI$'$ momentum scale employed is $a\bar q=\frac{2\pi}{32}\left(4{+}\frac{1}{4},0,0,4\right)$.}}
\label{fig2}
\end{minipage}
\end{center}
\end{figure}

It is interesting to plot the conversion factors for the cases used in simulations, that is,
$\mathcal{C}_{V_1(A_1)}$ and $\mathcal{C}_T$. For convenience we choose the coupling constant and the 
RI$'$ momentum scale to match the ensemble of twisted mass fermions employed in 
Ref.~\cite{Alexandrou:2016jqi}:
$g^2{=}3.077$, $a{=}0.082$fm, lattice size: $32^3\times64$ and $a\bar q{=}\frac{2\pi}{32} (\frac{n_t}{2}{+}\frac{1}{4},0,0,n_z)$, 
for $n_t{=}8$ and $n_z{=}4$ (the nucleon is boosted in the $z$ direction). The $\MSbar$ scale is set to $\bar\mu = 2{\rm GeV}$. The conversion 
factors are gauge dependent and we choose the Landau gauge which is mostly used in non-perturbative renormalization. 
Since the RI$'$ scale is given in lattice units, $a\bar q$, we also rescale the length of the Wilson line with the lattice spacing,
that is $z/a$. 
In Fig.~\ref{fig2} we plot the real (left panel) and imaginary (right panel) parts of $\mathcal{C}_{V_1(A_1)}$ and $\mathcal{C}_T$,
as a function of $z/a$. We remind the reader that the results of Eq.~(\ref{C_Gamma}) are 
only valid for $z{\neq} 0$. Thus, the points shown in the plot at $z{=}0$ (open symbols) have been extracted from
Ref.~\cite{Gracey:2000am}, and is a real function. In particular, it is exactly one for the scale and scheme 
independent vector and axial operators.

In the plot we allow $z$ to take all possible values up to half the lattice size, for both forward and backward directions of 
the Wilson line. One observes that the real part is symmetric with respect 
to $z{=}0$, while the imaginary part is antisymmetric. 
Additionally, for large values of $z$ the dependence of the conversion factor on the choice of operator becomes
milder. However, this behavior is not granted at higher loops, where a more pronounced dependence on the operator may arise.

\subsection{Lattice Regularization}
\label{resultsLR}

We now turn to the evaluation of the lattice-regularized bare Green's functions $\langle \psi\,{\cal O}_{\Gamma_1}\,\bar
\psi \rangle^{LR}$\,; this is
a far more complicated calculation, as compared to dimensional regularization, because the extraction of the divergences
is more delicate. 
The main task is to write lattice expressions in terms of continuum integrals, such as those in Appendix A, plus
additional terms
which are lattice integrals, independent of external momentum $q$; however, in contrast to the case of local operators,
these 
additional terms are expected to have a nontrivial dependence on $z$. Thus, the renormalized Green's functions stemming
from the lattice can be made to coincide with the continuum ones, shown in Appendix B, once an appropriate
$q$-independent (but $z$-dependent!) renormalization is applied.

The one-loop Feynman diagrams in LR are the same as in DR (Fig.~\ref{fig1}). Further, diagram 1 gives exactly the same
contribution as in DR; this was to be expected by the fact that the latter is finite as $\epsilon\to 0$, and
thus the limit $a\to 0$ can be taken right from the start, with no lattice corrections.

Once the loop momentum $p$ is rescaled to fit the boundaries of the Brillouin zone, $p\to p/a$, lattice divergences
manifest 
themselves as IR divergences in the external momentum $q\to 0$. We turn next to this issue, before giving our results in
LR.

\subsubsection{Isolation of IR divergences}

One of the most laborious tasks in the lattice computation is the extraction of IR divergences,
a process much more complex than in dimensional regularization. For demonstration purposes
we present the lattice counterpart of the integral discussed in DR (Eq.~(\ref{IntDR})), in order to point to
the necessity of applying novel techniques for its evaluation. 
In particular we developed a procedure for the isolation of IR divergences, somewhat in the spirit of 
the standard procedure of Kawai et al.~\cite{Kawai:1980ja}, in which one subtracts and adds 
to the original integrand its na\"\i ve Taylor expansion. In the present case,
the steps required are more complicated, and are presented below in a nutshell. 

A lattice expression analogous to Eq.~(\ref{IntDR}) is:
\be
I_{lat} \equiv \int \frac{dp^4}{(2\pi)^4}\,\frac{e^{-i\,n\,p_\mu}-1}{\sin{\left(\frac{p_\mu}{2}\right)}} \,\,
\frac{\sin{(p_\rho+a\,q_\rho)}}{\widehat{p}^{\,2}\,\widehat{(p+a \,q)}^2}\,,
\label{IntLR}
\ee
where $\hat{k}^2\equiv 4 \sum_\rho \sin^2(p_\rho/2)$, $\mu$ is the direction of the 
Wilson line, $\rho$ is in one of the four directions, and  $n\equiv z/a$; $I_{lat} $ appears
in the two `sail' diagrams. 
Below we present a schematic way of the process that we developed for the evaluation
of $I_{lat}$, which is based on a series of additions and subtractions. To avoid complicated
expressions we present the process in the Feynman gauge, even though in our calculation
we have kept a general gauge parameter $\beta$. 

We write: $I_{lat} = (I_{lat} - I_1) + (I_1-I_2) + (I_2-I_3) + I_3$, where:
\bea
I_{lat}&=& \int \frac{dp^4}{(2\pi)^4}\,\frac{e^{-i\,n\,p_\mu}-1}{\sin{\left(\frac{p_\mu}{2}\right)}} \,\,
\frac{\sin{(p_\rho+a\,q_\rho)}}{\widehat{p}^{\,2}\,\widehat{(p+a \,q)}^2} \nonumber\\[2ex]
&\Bigg\downarrow& \quad  I_{lat}-I_1 : {\rm Factorizable\ integral} \Rightarrow {\rm Explicit\ extraction\ of\ }a{\rm -dependence}
\nonumber\\[2ex]
I_1&=& \int \frac{dp^4}{(2\pi)^4}\,\frac{e^{-i\,n\,p_\mu}-1}{\sin{\left(\frac{p_\mu}{2}\right)}} \,\,
\left(\frac{\sin{(p_\rho+a\,q_\rho)}}{\widehat{p}^{\,2}\,\widehat{(p+a \,q)}^2} -
\frac{\sin{({\overline p}_\rho+a\,q_\rho)}}{\widehat{{\overline p}}^{\,2}\,\widehat{({\overline p}+a \,q)}^2}\right)
\label{IA}\\[2ex]
&\Bigg\downarrow& \quad I_1-I_2 : {\rm Naive\ } a\to 0 {\rm \ limit:\ } e^{-i\,n\,p_\mu}\to 0,\ a\,q\to 0 \Rightarrow 
{\rm \ Constant\ integral}\nonumber\\[2ex]
I_2&=& \int \frac{dp^4}{(2\pi)^4}\,\frac{e^{-i\,n\,p_\mu}-1}{p_\mu/2} \,\,
\left(\frac{p_\rho+a\,q_\rho}{p^2\,(p+a \,q)^2} -
\frac{{\overline p}_\rho+a\,q_\rho}{{\overline p}^{\,2}\,({\overline p}+a \,q)^2}\right)
\label{IB}\\[2ex]
&\Bigg\downarrow& \quad I_2-I_3 : {\rm Naive\ } a\to 0 {\rm \ limit:\ } e^{-i\,n\,p_\mu}\to 0,\ a\,q\to 0 \Rightarrow 
r{\rm -dependent\ integral}\nonumber\\[2ex]
I_3&=&\int_{p\le r} \frac{dp^4}{(2\pi)^4} \,\frac{e^{-i\,n\,p_\mu}-1}{p_\mu/2} 
\left( \frac{(p_\rho+a\,q_\rho)}{p^2\,(p+a \,q)^2} - \frac{(\overline p_\rho+a\,q_\rho)}{\overline p^{\,2}\,(\overline p+a \,q)^2}\right)
\nonumber \\[2ex]
&=& \int_{p\le r/a} \frac{dp^4}{(2\pi)^4} \frac{e^{-i\,z\,p_\mu}-1}{p_\mu/2}
 \left( \frac{(p_\rho+q_\rho)}{p^2\,(p+q)^2} - \frac{(\overline p_\rho+q_\rho)}{\overline p^{\,2}\,(\overline p+q)^2}\right).
\label{IC}
\eea
Here, $\overline p$ is the loop momentum $p$ with its $\mu$-component set to 0.
In Eq.~(\ref{IC}) $r$ represents the radius of a sphere over which we integrate, and has an arbitrary value. 
Each of the two terms in Eq.~(\ref{IC}) can be calculated by integration in spherical coordinates, leading to the same
expressions as those found in DR in terms of Bessel functions, plus $q$-independent terms.
Naturally, the integrals over the two terms of Eq.~(\ref{IC}) depend on the parameter $r$. However, once all parts of
$I_{lat}$ are combined, the final result is  
independent of $r$; this cancellation is highly nontrivial and it provides an important cross-check of 
our calculation. 

\subsubsection{Multiplicative Renormalization and Mixing}
\label{MRaM}

Despite the complexity of the bare Green's functions, their 
difference in DR and LR is necessarily polynomial in the external momentum (of degree 0 in this case), which leads to a
prescription for extracting $Z_\mathcal{O}^{LR,\MSbar} $ without an intermediate (e.g., RI$'$-type) scheme. 

By analogy with closed Wilson loops~\cite{Dotsenko:1979wR} in regularizations other than DR, we find a linear
divergence also for Wilson line operators in LR; it is proportional to $|z|/a$ and arises from the 
tadpole diagram (d4), with a proportionality coefficient which depends solely on the choice of the gluon action. The
exact term that leads to such a divergence is given by:
\bea
\int_{-\pi}^\pi \frac{d p^4}{(2\pi)^4}\,S_g(p)_{\mu\mu} \, 
\frac{\sin^2{\left(\frac{z}{a}\,\frac{p_\mu}{2}\right)}}{\sin^2(p_\mu/2)}
&=& \int_{-\pi}^\pi \frac{d p^3}{(2\pi)^3}\,S_g(\overline p)_{\mu\mu} \cdot \int_{-\pi}^\pi \frac{d p_\mu}{(2\pi)}\,
\frac{\sin^2{\left(\frac{z}{a}\,\frac{p_\mu}{2}\right)}}{p_\mu^2/4} + {\cal O}(a^0,\,\log a) \nonumber \\
 &=& \int_{-\pi}^\pi \frac{d p^3}{(2\pi)^3}\,S_g(\overline p)_{\mu\mu} \cdot \left(-\frac{2}{\pi^2} + \frac{|z|}{a} \right)  + {\cal O}(a^0,\,\log a) \,,
\label{LinearDiv}
\eea
where $S_g(p)_{\mu\mu}$ is the diagonal matrix element of the gluon propagator in the direction of the
Wilson line ($\mu$) and $\overline p$ is the loop momentum $p$ with its $\mu$-component set to 0.

Just as with other contributions to the bare Green's function, the linear divergence is the same -- at one-loop level -- for
all operator insertions. In a resummation of all orders in perturbation theory, the powers of $|z|/a$ are expected to
combine into an exponential of the form~\cite{Dotsenko:1979wR}:
\begin{equation}
\Lambda_{\Gamma} = e^{-c \, |z|/a} \, {\tilde\Lambda}_\Gamma \,,
\label{LambdaTilde}
\end{equation}
where $c$ is given by Eq.~(\ref{LinearDiv}) to one loop, and ${\tilde\Lambda}_\Gamma$ is related to
$\Lambda_\Gamma^\MSbar$ by a further renormalization factor which is at most logarithmically divergent with $a$.
Based on arguments from heavy quark effective theory, additional contributions may appear in 
the exponent~\cite{Sommer:2015hea}. This will be discussed in subsection~\ref{sub33}.

To one loop, we find the following form for the difference between the bare lattice Green's functions and the
$\MSbar$-renormalized ones: 
\bea
\langle \psi\,{\cal O}_{\Gamma}\,\bar \psi \rangle^{DR, \,\MSbar}
-\langle \psi\,{\cal O}_{\Gamma}\,\bar \psi \rangle^{LR} = 
\frac{g^2\,C_f}{16\,\pi^2}\, e^{i\,q_\mu z}\, \hspace*{-0.4cm}&\Bigg[&\hspace*{-0.3cm}\Gamma \Big(\alpha_1 + \alpha_2\, \beta + \alpha_3\,\frac{|z|}{a}  
+ \log \left(a^2 \bar\mu^2\right)\left(4-\beta\right) \Big)\nonumber\\
&&\hspace*{-0.3cm} +\left(\Gamma\cdot\gamma_\mu + \gamma_\mu\cdot\Gamma \right)\,\Big(\alpha_4 + \alpha_5\,c_{\rm SW}\Big) \Bigg]\,.
\label{diffDRLR}
\eea
Using Eq.~(\ref{diffDRLR}) together with  Eq.~(\ref{eq2}) one can extract the multiplicative renormalization and mixing coefficients
in the $\MSbar$-scheme and LR.
In our calculation the coefficient $\alpha_3$ is negative, in accordance with Eq.~(\ref{LambdaTilde}).

In Eq.~(\ref{diffDRLR}) all coefficients $\alpha_i$ depend on the Symanzik parameters, except for $\alpha_2$ which has a
numerical value $\alpha_2 = 3.792$. This value 
was expected, as all gauge dependence must disappear in the $\MSbar$ scheme for gauge invariant operators: Indeed, this
term will cancel against a similar term in 
$Z^{LR, \,\MSbar}_\psi$ in Eq.~(\ref{eq2}). The latter has been computed using the same set-up in a previous work
\cite{Alexandrou:2012mt} and has the general form:
\be
Z^{LR, \,\MSbar}_\psi = 1 + \frac{g^2\,C_f}{16\,\pi^2}\, \Bigg[e^\psi_1 +3.792\,\beta + 
e^\psi_2\,c_{\rm SW} + e^\psi_3\,c_{\rm SW}^2 + (1 - \beta) \log\left(a^2 \bar\mu^2\right)\Bigg]\,.
\label{Zpsi}
\ee
A few interesting properties of Eq.~(\ref{diffDRLR}) can be pointed out: 
The contribution $\left(\Gamma\cdot\gamma_\mu {+} \gamma_\mu\cdot\Gamma \right)$ indicates mixing between operators
of equal dimension, which is finite and appears in the lattice regularization. 
To one-loop level we find that the mixing matrix is symmetric, and thus, its
eigenvalues are the same as those of the addition and difference of the
operators which mix which each other. We would also like to note that
mixing of different origin has been discussed in Ref.~\cite{Radyushkin:2016hsy} using the
parton virtuality distribution functions (VDFs) formalism.
Moreover, this combination vanishes for certain choices of the Dirac structure $\Gamma$ in the operator. 
For the operators $P$, $V_\nu$ ($\nu\ne\mu$), $A_\mu$, $T_{\mu\nu}$ ($\nu\ne\mu$) the combination
$\left(\Gamma\cdot\gamma_\mu + \gamma_\mu\cdot\Gamma \right)$  
is zero and only a multiplicative renormalization is required. This has significant impact in the non-perturbative calculation of the
unpolarized quasi-PDFs, as there is a mixing with a twist-3 scalar operator~\cite{Jaffe:1991ra}.
Such a mixing must be eliminated using a proper renormalization prescription, ideally non-perturbatively~\cite{Alexandrou:2017huk}.

To one-loop level, the diagonal elements of the mixing matrix (multiplicative renormalization) are the same for all
operators under study, and through Eq.~(\ref{eq2}) one obtains: 
\be
Z_\Gamma^{LR,\MSbar}  =  1 +  \frac{g^2\,C_f}{16\,\pi^2}\, \left( e_1 + e_2 \,\frac{|z|}{a}  
+e_3\,c_{\rm SW} + e_4\,c_{\rm SW}^2
-3 \log \left(a^2  \bar\mu^2\right) \right),
\label{Zmult}
\ee
where the coefficients $e_1 - e_4$ are given in Table~\ref{tab2}, for the Wilson, tree-level Symanzik and Iwasaki improved actions. 
Results on other gluonic actions can be provided upon request. It is worth mentioning that the presence of $c_{\rm SW}$
in $Z_\Gamma^{LR,\MSbar}$ is inherited from $Z_\psi$. 
As expected, $Z_\Gamma^{LR,\MSbar}$ is gauge independent, and the cancelation of the gauge dependence was numerically 
confirmed up to ${\cal O}(10^{-5})$. This gives an estimate on the accuracy of the numerical loop integrations. The systematic
error stemming from our integration procedure is taken into account by reporting only 6 significant digits.
Similar to $Z_\Gamma^{LR,\MSbar}$, the nonvanishing mixing coefficients are operator independent and have the general form:
\be
Z_{12}^{LR,\MSbar}  = Z_{21}^{LR,\MSbar}  =   0 +  \frac{g^2\,C_f}{16\,\pi^2}\, \left( e_5+ e_6 \,c_{\rm SW} \right),
\label{Zmix}
\ee
where $Z_{ij}^{LR,\MSbar}$ ($i\ne j$) is nonzero only for the operator pairs: $\{S,\, V_1\}$, $\{A_2,\, T_{34}\}$, $\{A_3,\,
T_{42}\}$, $\{A_4,\, T_{23}\}$.
The values of the coefficients $e_5$ and $e_6$ for Wilson, tree-level Symanzik and Iwasaki gluons 
are shown in the last two columns of Table~\ref{tab2}. 
Unlike the diagonal elements, the dependence of $Z_{12}^{LR,\MSbar}$ on the clover parameter is genuinely extracted 
from the Green's functions of the operators. Given that the strength of mixing depends on the value of $c_{\rm SW}$, 
one may consider increasing $c_{\rm SW}$ from 0 to $-e_5/e_6$ in order to suppress mixing: 
for example, choosing: $c_{\rm SW}=0,\, 1,\, 1.5$ for the Iwasaki action, as shown in Table~\ref{tab2}, the mixing coefficient is 
${\cal O}(g^2{\times}10^{-1}),\,{\cal O}(g^2{\times} 10^{-2}),\,{\cal O}(g^2{\times} 10^{-3})$, respectively,
and completely vanishes (to one loop) at $c_{\rm SW}=-e_5/e_6$. This information 
allows one to tune the clover parameter in order to eliminate mixing at one loop.

\begin{table}[h]
\begin{center}
\begin{tabular}{lcccccc}
\hline
\hline\\[-2ex]
 Action  &  $e_1$ &  $e_2$  &  $e_3$  & $e_4$  & $e_5$ &  $e_6$ \\[0.5ex]
\hline\\[-2ex]
Wilson				&24.3063	&-19.9548	&-2.24887	&-1.39727	&14.4499	&-8.28467		\\
TL Symanzik	\,\,\,	&19.8442	&-17.2937	&-2.01543	&-1.24220	&12.7558	&-7.67356 		\\
Iwasaki				&12.5576	&-12.9781	&-1.60101	&-0.97321	&9.93653	&-6.52764		\\
\hline
\hline
\end{tabular}
\vspace*{-0.4cm}
\begin{center}
\begin{minipage}{15cm}
\hspace*{3cm}
\caption{Numerical values of the coefficients $e_1$ - $e_4$ of the multiplicative renormalization functions and $e_5$ - $e_6$ of the 
mixing coefficients for Wilson, tree-level (TL) Symanzik and Iwasaki gluon actions.}
\label{tab2}
\end{minipage}
\end{center}
\end{center}
\end{table}

\subsubsection{Renormalized Green's functions}

We now present the renormalized Green's functions and discuss some of their important features which will allow us to set up 
a procedure for estimating the linear divergence using non-perturbative data. This procedure will be explained in the next Subsection.

The complete expressions for the renormalized Green's functions are listed in Eqs.~(\ref{SP_R})-(\ref{T2Tp2_R}) of Appendix B
for general values of the momentum, length of Wilson line, and gauge fixing parameter. As the renormalized Green's functions
do not depend on the regularization choice, their expressions do not contain $c_{\rm SW}$ and the lattice spacing.
For ease of notation, Eqs.~(\ref{SP_R})-(\ref{T2Tp2_R}) are expressed in terms of the integrals $F_1$ - $F_5$ and $G_1$ - $G_5$
of Appendix A. In the present Subsection we focus on the simpler case in which the external momentum, $q$, is in the same
direction as the Wilson line, $\mu$\,; we will thus denote $q_\mu$ simply by $q$. In this case the renormalized Green's
functions are multiples of their tree-level values.
In particular, they can be cast into the form:
\be
\displaystyle
 \Lambda^{\rm 1-loop}_\Gamma\Bigg{|}_{(q,0,0,0)} \hspace*{-0.5cm}= \Lambda^{\rm tree}_\Gamma\,
 \left(\frac{\bar{\mu}^2}{q^2}\right)^{(4-\beta) g^2\,C_f/(16\,\pi^2)}
\Bigg[1 +  \frac{g^2\,C_f}{16\,\pi^2}  F_\Gamma(qz) \Bigg]\,,
\label{LL}
\ee
where $F_\Gamma$ are complicated complex functions of $qz$, satisfying $F_\Gamma(-x) =  F_\Gamma^\dagger(x) $.
In standard fashion, we have exponentiated the one-loop result for the 
$\bar\mu$-dependence, thus putting into evidence the anomalous dimension
of the Green's functions. Once again we note that this scale dependence does not vary with the length of the Wilson
line; this is in agreement with older calculations, regarding closed Wilson loops~\cite{Dotsenko:1979wR}.

The form of Eq.~(\ref{LL}) is in accordance with the features of the physical matrix elements of Wilson line operators
computed in numerical simulations.  The requirement that renormalization functions be $q$-independent implies that the
dependence on $aq$ in the bare Green's functions can be predicted from the start, as it has to match the $\bar\mu$ 
dependence of the renormalized Green's functions. Below we provide the expressions for $F_\Gamma$:
\bea
F_{S(P)}(qz) \hspace*{-.2cm}&=&\hspace*{-.2cm}  
8 + 4\gamma_E -\log(16) -2  |q| |z|( {F_4}-2 {F_5}) +8 {F_2} +(\beta +2) \log \left( q^2 z^2\right)\nonumber\\
     &+& \hspace*{-.2cm}\beta  \Big[-3 + 2 \gamma_E - \log(4)- 
    | q| |z|{F_4}+2  {F_1}+ q^2 \left(z^2
    \left(\frac{ {F_1-F_2}}{2}- {F_3}\right)+4  G_3\right)\Big]\nonumber\\
&+& \hspace*{-.2cm} i  q\Bigg\{4 (z
    ( {F_1-F_2}- {F_3})- {G_1}) -\beta  \Big[| q| (  z |z| {F_5}+2
    ( {G_4}- 2{G_5}))+2(G_2-G_1)\Big]\Bigg\}
    \label{F_SP}
\eea
\bea
F_{V_1(A_1)}(qz) \hspace*{-.2cm}&=&\hspace*{-.2cm}    
 8 + 4\gamma_E -\log(16)  +4 |q| |z|( {F_4}-2 {F_5}) -4  {F_2}  +(\beta +2) \log
    \left( q^2 z^2\right)\nonumber\\
&+& \hspace*{-.2cm}  \beta  \Big[-3 + 2 \gamma_E - \log(4)+2  
    | q| |z|{F_4}+2  {F_1}+ q^2 \left(z^2
    \left(\frac{ {F_1-F_2}}{2}- {F_3}\right)+4  {G_3}\right)\Big]\nonumber\\
&+& \hspace*{-.2cm} i  q\Bigg\{{-}4 (2 z
    ( {F_1-F_2}- {F_3})+ {G_1}) -\beta \Big[ | q| ( 
    z |z|{F_5}+2 ( {G_4}- 2{G_5}))+2(G_2-G_1)\Big]\Bigg\} 
\eea
\vspace*{0.25cm}
\bea
F_{V_\nu(A_\nu)}(qz) \hspace*{-.2cm}&=&\hspace*{-.2cm}   
8 + 4\gamma_E -\log(16)-4  {F_2}  +(\beta +2) \log \left( q^2 z^2\right)\nonumber\\
&+& \hspace*{-.2cm}\beta
    \Big[-3 + 2 \gamma_E - \log(4)+2  {F_1}+ q^2 \left(z^2
    \left(\frac{ {F_1-F_2}}{2}- {F_3}\right)+4  {G_3}\right)\Big]\nonumber\\
&+& \hspace*{-.2cm} i  q \Bigg\{-4  {G_1}- \beta  \Big[| q| (  z |z|{F_5}+2
    ( {G_4}- 2{G_5}))+2(G_2 -G_1)\Big]\Bigg\} \qquad\qquad (\nu\ne 1) \hspace*{3cm}
\eea
\vspace*{0.25cm}
\bea
F_{T_{1\nu}}(qz) \hspace*{-.2cm}&=&\hspace*{-.2cm}  
8 + 4\gamma_E -\log(16)+2 |q| |z|({F_4} -2 {F_5})-8{F_2}+(\beta +2) \log
    \left( q^2 z^2\right) \nonumber\\
&+& \hspace*{-.2cm}\beta  \Big[-3 + 2 \gamma_E - \log(4) + 
    | q| |z|{F_4}+2 {F_1}+ q^2 \left(z^2
    \left(\frac{ {F_1-F_2}}{2}- {F_3}\right)+4  {G_3}\right)\Big]\nonumber\\
&+& \hspace*{-.2cm} i  q\Bigg\{-4  (z
    ( {F_1-F_2}- {F_3})+ {G_1}) -\beta  \Big[ | q| (  z |z|{F_5}+2
    ( {G_4}- 2{G_5}))+2 (G_2- G_1) \Big]\Bigg\}\\[5ex]
F_{T_{\nu\rho}}(qz) \hspace*{-.2cm}&=&\hspace*{-.2cm}   F_{T_{1\nu}} \qquad (\nu\ne 1,\ \rho\ne 1).
\label{F_T}
\eea

\begin{figure}[h!]
\centerline{\includegraphics[scale=.285,angle=-90]{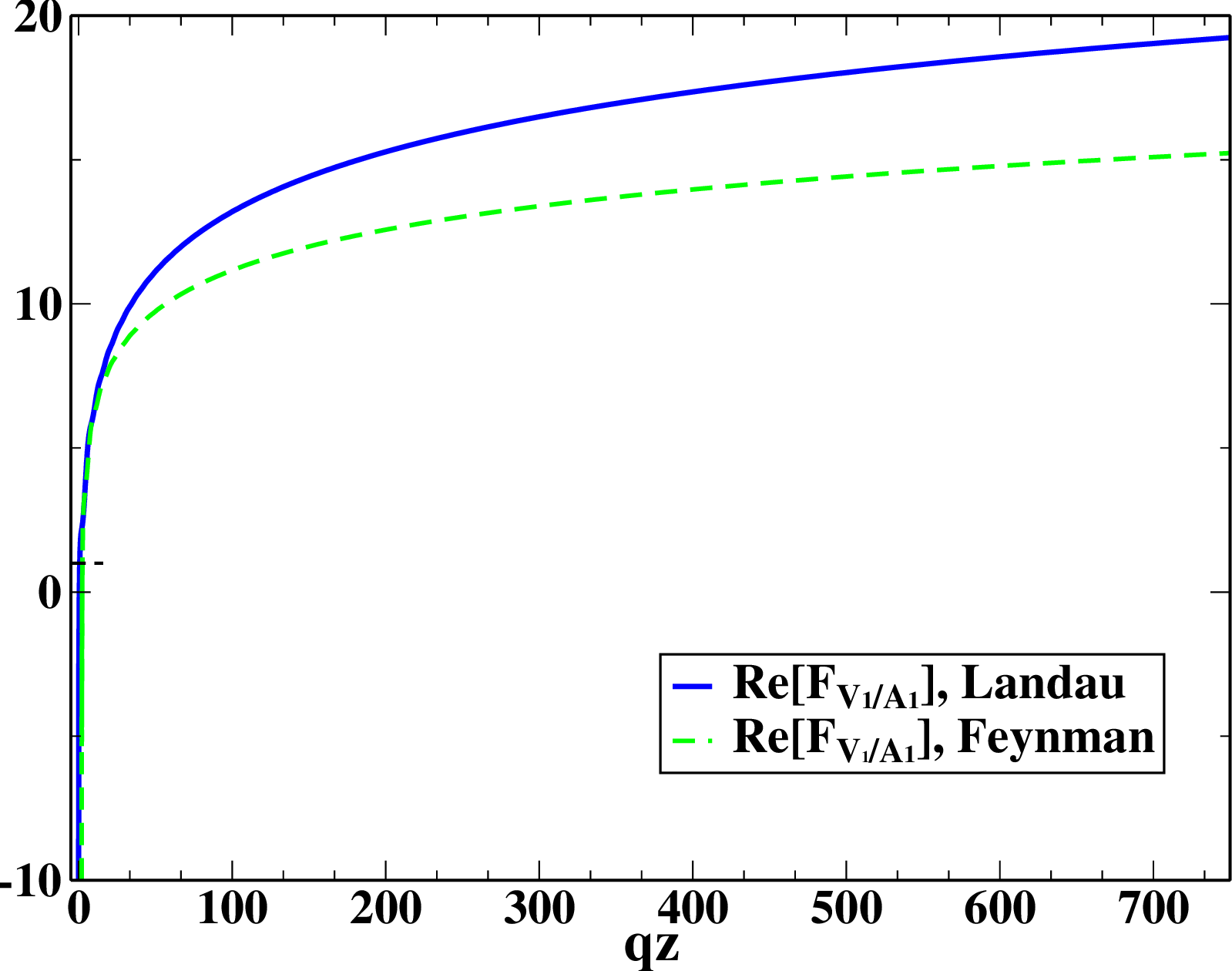}\quad
\includegraphics[scale=.285,angle=-90]{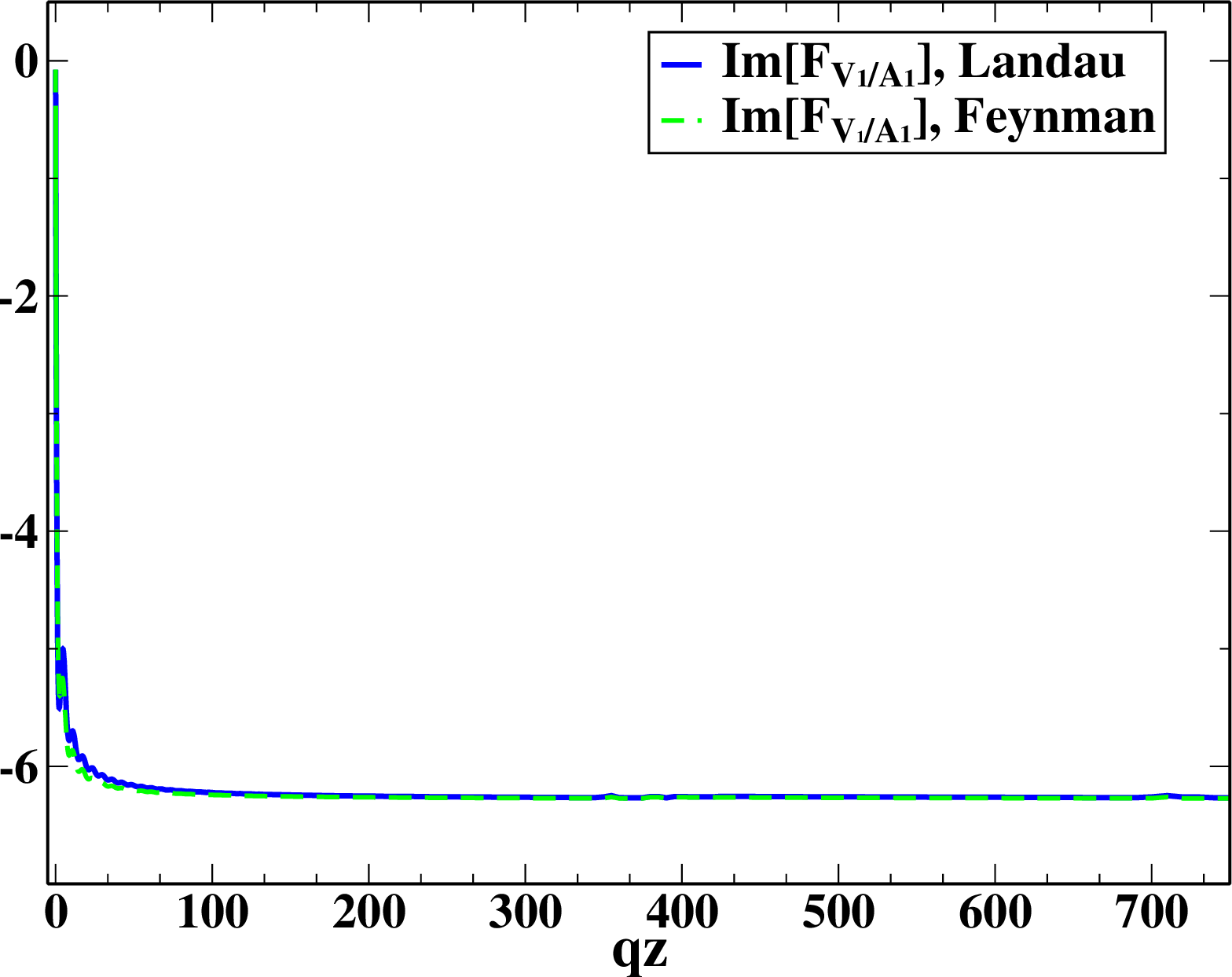}}
\vspace*{-0.3cm}
\begin{center}
\begin{minipage}{15cm}
\hspace*{3cm}
\caption{\small{Real (left panel) and imaginary (right panel)} parts for the function $F_{V_1(A_1)}$ versus
$qz$. Solid and dashed lines correspond to the Landau and Feynman gauge, respectively.}
\label{fig3}
\end{minipage}
\end{center}
\end{figure}

In Appendix B we present the $\MSbar$-renormalized Green's functions for each Wilson line operator, and for general RI$'$
renormalization scale 4-vector $\bar q_\nu$\,. 
For demonstration purposes, in Fig.~\ref{fig3} we plot $F_{V_1(A_1)}$ in the Landau and Feynman gauge as a function
of the dimensionless quantity $qz$. We find that the qualitative behavior of $F_{V_1(A_1)}$ 
is similar regardless of the choice for $\beta$. It is interesting to see the limit $q \to 0$ for $F_{V_1(A_1)}(qz)$, 
which, in fact, coincides with the limit $z\to 0$. As mentioned earlier, the latter is expected to be singular, due to
the emergence of contact terms. Thus, as $q\to 0$, we find that 
$F_{V_1(A_1)}(qz)$ is a real function independent of the gauge and equal to:
\begin{equation}
\lim_{q\to 0}  F_{V_1(A_1)}(qz) = 3 \left(1 + \gamma_E + \log\left(\frac{qz}{2}\right) \right)\,.
\end{equation}

\newpage
\subsection{Non-perturbative elimination of the linear divergence} 
\label{sub33}
 
As shown in Subsection~\ref{MRaM}, there is a linear divergence in the lattice-regularized Wilson line operator, which
requires a careful removal before the continuum limit can be reached in the non-perturbative matrix elements. One way to
eliminate this divergence is to use the estimate of the  
one-loop coefficient $e_2$ of Eq.~(\ref{Zmult}) and subtract it from the non-perturbative matrix elements. We have computed $e_2$ 
for a large class of gluon actions\footnote{We recall that the value of $e_2$ does not depend on the fermion action.}, and presented in
this paper the results for Wilson, TL Symanzik and Iwasaki actions.
However, this subtraction only partially removes the divergence, as higher orders still remain and they will dominate in the
$a\to 0$ limit. In general, power divergences are features in which deviations from perturbation theory are most
pronounced. Thus,
it is preferable to develop a non-perturbative method to extract the linear divergence. 

A method suggested recently~\cite{Ishikawa:2016znu} 
is to use the static potential in order to eliminate non-perturbatively the exponential with the linear divergence,
cf. Eq.~(\ref{LambdaTilde}). Here we propose an alternative 
methodology which is based on using bare matrix elements of the Wilson line operators from numerical simulations,
denoted by $q(P_3,z)$ in Ref.~\cite{Alexandrou:2016jqi}. We consider the case where
there is no mixing between operators and we focus on the helicity (axial) and transversity (tensor), indicated by
$\Delta h(P_3,z)$ and $\delta h(P_3,z)$, respectively,  
in Ref.~\cite{Alexandrou:2016jqi}. We note that both helicity and transversity have a Dirac index in the direction of the Wilson
line, which exhibits no mixing. In the non-perturbative calculations of the matrix elements, the nucleon is boosted by
momentum $P_3$ which is in the same direction as the Wilson line, simplifying the calculation. Based on the arguments
presented in the previous Subsection (Eq.~(\ref{LL})) we expect that the renormalized matrix elements can depend on $z$
only through the dimensionless quantity 
$P_3 z$. Furthermore, the dependence on the scale $\bar\mu$ in the renormalized matrix element is well defined and involves the anomalous 
dimension ($\gamma_\Gamma$) of the operator: $q^R(P_3 z, P_3/\bar\mu)\propto {\bar\mu}^{-2\gamma_\Gamma}$, which is
matched by the $\bar\mu$ dependence in the renormalization function.  Thus, 
\be
q^R(P_3 z, P_3/\bar\mu) = (P_3/\bar\mu)^{2\gamma_\Gamma} \cdot {\tilde q}^R(P_3 z).
\ee 
Similarly, the renormalization function
$Z_\Gamma^{LR,\MSbar}(a\bar\mu, z/a)$, given its expected $\bar\mu$-dependence, will factorize as: 
\be
Z_\Gamma^{LR,\MSbar}(a\bar\mu, z/a) = {\tilde Z}_\Gamma(a\bar\mu)\cdot \hat Z(z/a).
\ee 
The factor ${\tilde Z}_\Gamma(a\bar\mu)$ is not simply $(a\bar\mu)^{2\gamma_\Gamma}$, due to finite lattice corrections;
in the one-loop case, these lattice corrections are the terms containing $e_1$\,, $e_3$\,, $e_4$ in
Eq.~(\ref{Zmult}). The factor $\hat Z(z/a)$ originates exclusively from tadpole diagrams, such as d4 of the one-loop
case. Based on the behavior of tadpole diagrams, one can see that the one-loop contribution proportional to $e_2$
in Eq.~(\ref{Zmult}) will exponentiate upon considering higher powers
of $g$, leading to: 
\be \hat Z(z/a) = e^{-\delta m\,|z|/a}, \qquad \delta m = - \frac{g^2\,C_f}{16\,\pi^2}\, e_2 + {\cal O}(g^4). \ee
This behavior is entirely consistent with the exponential behavior $\exp(-\delta m\,|z|/a)$ proven in
Ref.~\cite{Dotsenko:1979wR} for closed Wilson loops. We note that the proof in Ref.~\cite{Dotsenko:1979wR} holds for any
regularization in which $z/a$ terms may arise.

We can thus write the ratio of the bare matrix elements for different values of $P_3$ and $z$ as:
\be
\displaystyle
\frac{q(P_3, z)}{q(P'_3, z')} = 
\frac{Z_\Gamma^{LR,\MSbar}(a\bar\mu, z/a) \cdot q^R(P_3 z, P_3/\bar\mu) }
{Z_\Gamma^{LR,\MSbar}(a\bar\mu, z'/a) \cdot q^R(P'_3 z', P'_3/\bar\mu) } =
\frac{e^{-\delta m\,|z|/a} \, \tilde Z_\Gamma(a\bar\mu) \, \left({\displaystyle \frac{P_3}{\bar\mu}}\right)^{2\gamma_\Gamma} \, \tilde q^R(P_3 z) }
{e^{-\delta m\,|z'|/a} \, \tilde Z_\Gamma(a\bar\mu) \, \left({\displaystyle \frac{P'_3}{\bar\mu}}\right)^{2\gamma_\Gamma} \,
  \tilde q^R(P'_3 z')  } \,,
\label{ratio}
\ee
where the one-loop anomalous dimension is $\gamma_\Gamma =  -3 g^2 C_f/(16\pi^2)$ for all operator insertions. The anomalous dimension
of the fermion field is not relevant in this discussion as the non-perturbative matrix elements are between physical (nucleon) states.
In the above ratio, one may choose $P_3,\,P'_3,\,z,\,z'$ such that $P_3 z = P'_3 z'$, which simplifies the ratio considerably:
\be
\frac{q(P_3, z)}{q(P'_3, z')} = 
e^{-\delta m\,(|z|-|z'|)/a} \, \left(\frac{P_3}{P'_3}\right)^{-6 g^2 C_f / (16\pi^2) }\,.
\label{RR}
\ee
Thus, by forming the ratio $q(P_3, z)/q(P'_3, z')$ from non-perturbative data, and
by choosing several combinations of $P_3 z = P'_3 z'$, one can fit to extract the coefficient of the linear 
divergence, $\delta m$.

We note that there are non-perturbative arguments~\cite{Sommer:2015hea} to suggest that a further finite, dimensionful scale may
appear multiplying $z$ in the exponential term: $\exp(-\delta m\,|z|/a - c|z|)$. In this case, Eq.~(\ref{RR}) takes the
form:
\be
\frac{q(P_3, z)}{q(P'_3, z')} = 
e^{-\big((\delta m/a)+ c\big)\,(|z|-|z'|)} \, \left(\frac{P_3}{P'_3}\right)^{-6 g^2 C_f / (16\pi^2) }\,.
\label{RRcbar}
\ee
Thus, we may still deduce the value of the quantity $(\delta m/a)+ c$. Extracting the values of $\delta m$ and $c$ separately
would require utilizing simulation data from two or more values
of the lattice spacing $a$; however, this procedure will be hampered by the very dependence of $\delta m$ on the coupling
constant, as the lattice spacing is varied. We will not pursue further the non-perturbative evaluation of $\delta m$ and 
$c$, since it is beyond the scope of this paper. 

The ratio of the left-hand-side of Eq.~(\ref{RRcbar}) can be investigated for the helicity and transversity, 
which do not exhibit mixing. Since the right-hand side of Eq.~(\ref{RRcbar}) is independent 
of the operator insertion, one expects the same value for the exponential coefficient, up to lattice artifacts. 
We have tested this method with the data of ETMC presented in Ref.~\cite{Alexandrou:2016jqi}, with encouraging results: 
\begin{itemize}
\item The ratio $q(P_3, z)/q(P'_3, z') $ was found to be real for the helicity and transversity
 as expected from Eq.~(\ref{RRcbar}), despite the fact that the matrix elements themselves are complex; 
\item The analogous ratio for the unpolarized operator, which mixes with the scalar, leads to a nonzero imaginary part;
\item The extracted value for the coefficient $(\delta m/a)+c$, using different combinations of $P_3 z$, is consistent within statistical 
accuracy; 
\item Both helicity and transversity give very similar estimates for $(\delta m/a)+c$.
\end{itemize}
\begin{figure}[h]
\centerline{\includegraphics[scale=.375,angle=-90]{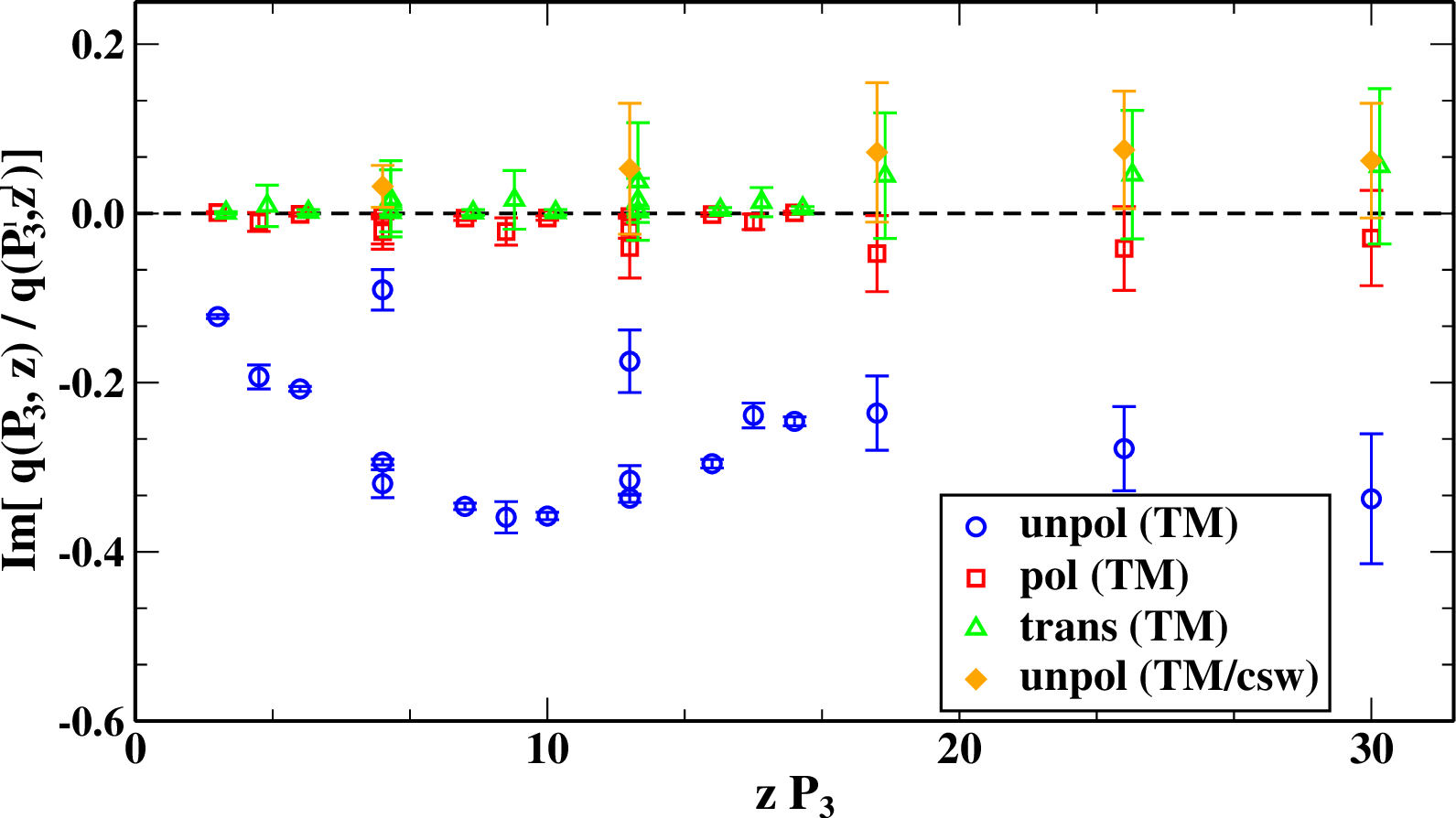}}
\vspace*{-0.3cm}
\begin{center}
\begin{minipage}{15cm}
\hspace*{3cm}
\caption{\small{The imaginary part of the ratio of Eq.~(\ref{RR}) for simulation data at $m_\pi{=}375$\,MeV
using Twisted Mass fermions from Ref.~\cite{Alexandrou:2016jqi} (open symbols), as well as preliminary
data at $m_\pi{=}130$\,MeV using Twisted Mass clover-improved fermions (filled symbols). 
The unpolarized (blue circles, filled orange diamonds), polarized (red squares) and transversity (green triangles) 
quasi-PDFs are presented.}}
\label{fig4}
\end{minipage}
\end{center}
\end{figure}

Indeed, Fig.~\ref{fig4} shows that the unpolarized case from Ref.~\cite{Alexandrou:2016jqi} (blue circles)
has a nonzero imaginary part, while the imaginary parts of the ratios for helicity and for transversity on the
same ensemble are compatible with zero. The simulation data of Ref.~\cite{Alexandrou:2016jqi} regard 
twisted mass fermions at a pion mass of $m_\pi{=}375$\,MeV, and Iwasaki gluons. 
For these action parameters and $\beta{=}1.95$, our perturbative results of Eq.~(\ref{Zmix}) show a
significant mixing, that is $(6/\beta)\cdot C_f/(16\pi^2) \cdot \left(9.93653\right) {\sim} 0.26$.
This is confirmed by the data of Fig.~\ref{fig4}.  It is also very interesting to test the left-hand-side of Eq.~(\ref{RRcbar}) on an ensemble in
which the mixing is expected to be very small. ETMC has preliminary data at the physical point for twisted mass fermions ($\beta{=}2.1$)
including a clover term $c_{\rm SW}{\sim}1.57$, and Iwasaki gluons. Thus, according to the results in Eq.~(\ref{Zmix}) and the
numerical values presented in Table~\ref{tab2}, the one-loop mixing is reduced by two orders of magnitude as  $(6/\beta)\cdot C_f/(16\pi^2) \cdot \left(9.93653 -6.52764\, c_{\rm SW}\right) {\sim} \,{-}0.00752345$.
The fact that the mixing is insignificant in the presence of a clover term with $c_{\rm SW}{\sim}1.57$, is also confirmed by the
simulation data (orange filled diamonds), as shown in Fig.~\ref{fig4}. We note that even though the statistical errors are still large,  
the imaginary part of  Eq.~(\ref{RRcbar}) is zero within error bars.

The above demonstration using simulation data stresses the importance of eliminating the mixing
between the vector and scalar operators, prior to applying the matching procedure LaMET (Large-Momentum Effective Field
Theory) to the physical PDFs. This can be achieved 
by computing the matrix elements for both the scalar and vector operators. One may use our perturbative
results for the multiplicative renormalization and the mixing coefficients to disentangle the two operators
and extract the renormalized unpolarized quasi-PDFs. Once this is done, one can apply the ratio procedure of
Eqs.~(\ref{ratio})-(\ref{RR}) to these operators as well.

 \section{Summary - Future Work}
 
In this paper we have presented the one-loop perturbative calculation of the renormalization functions for
operators including a straight Wilson line. Results of this work may be used to renormalize, to one-loop
in perturbation theory, matrix elements for quasi-PDFs. Two main features of the quasi-PDFs in lattice regularization
are demonstrated in this work: finite mixing and a linear divergence with respect to 
the regulator. We computed the one-loop Green's functions
both in dimensional (DR) and lattice (LR) regularizations, which allows one to extract the LR renormalization 
functions in the $\MSbar$-scheme directly, without an intermediate RI$'$ scheme. Using the Green's 
functions in DR we computed the conversion factor between an RI$'$-type (Eq.~(\ref{eq6})) scheme and 
$\MSbar$. These expressions are necessary to bring non-perturbative estimates of the renormalization
functions to the $\MSbar$-scheme. 

We demonstrate for the first time that certain Wilson line operators exhibit mixing within the lattice regularization. 
This has great impact on the unpolarized quasi-PDFs that mix with a twist-3~\cite{Jaffe:1991ra} scalar operator. 
Thus, before matching to the physical PDFs one must eliminate the mixing, which would require computation of the 
matrix elements for the scalar operator.  However, our results indicate that the presence of a clover term in the fermion 
action highly suppresses the mixing, as can be read from Eq.~(\ref{Zmix}). This has been tested on nucleon 
matrix elements of the unpolarized quasi-PDFs using twisted mass fermions, with and without clover improvement 
(Fig.~\ref{fig4}), confirming what is expected from our perturbative results. This may also be 
an indication that higher loop contributions are suppressed. Furthermore, the mixing discussed for the unpolarized
case vanishes if one uses a Dirac structure perpendicular to the Wilson line direction, as explained in the main
text. For example, choosing the $\gamma$-matrix in the temporal direction is important for a faster
convergence to the physical PDFs, as discussed in Ref.~\cite{Radyushkin:2016hsy}.

A feature of the quasi-PDFs that requires special treatment is the linear divergence that complicates
taking the continuum limit. We have computed the one-loop coefficient of the linear divergence for a variety
of gluonic actions, which can be subtracted from nucleon matrix elements of the quasi-PDFs. Such a
divergence is expected to resum to an exponential, just as in the case of Wilson loop operators~\cite{Dotsenko:1979wR}.
Using the form of the renormalized Green's function of Eq.~(\ref{LL})
we propose a technique to extract the coefficient of the linear divergence in Subsection~\ref{sub33} 
from non-perturbative data.

Using the renormalization pattern exposed in  this work we have developed an appropriate non-perturbative 
renormalization prescription for the unpolarized, helicity and transversity quasi-PDFs. Such a scheme will 
extract both the renormalization functions and linear divergence at once. For the unpolarized case, the 
mixing with the twist-3 scalar operator is also addressed. This will be presented in a follow-up publication
\cite{Alexandrou:2017huk}.

A natural continuation of this project is the addition of smearing to the fermionic part of the action and/or to the
gauge links of the Wilson line operator. This is important, as modern simulations employ such smearing techniques 
(e.g., stout and Hypercubic (HYP)) that suppress the power divergence and bring the renormalization functions closer to their
tree-level values. Smearing the operator under study alters its renormalization functions, thus, the same smearing 
must be employed in the renormalization process. 
In these cases, the additional contributions to the renormalization functions due to smearing are more convergent, 
and thus the perturbative extraction of singularities is simpler. Nevertheless, the smearing of the gauge links results 
in a huge increase of the number of terms in the vertices and, thus, the computation of the diagrams becomes very challenging. 

An extension of this calculation that we intend to pursue, is the evaluation of lattice artifacts to one loop and to all orders
in the lattice spacing, ${\cal O}(a^\infty,\,g^2)$, as developed in Ref.~\cite{Gockeler:2010yr,Constantinou:2013ada}. This has been 
successfully applied for local and one-derivative fermion operators eliminating lattice artifacts from non-perturbative
estimates~\cite{Constantinou:2014fka,Alexandrou:2015sea}. For operators with a long Wilson line 
($z {>>} a$) the lattice artifacts are likely to be more prominent, and therefore, such a calculation will be
extremely beneficial in the non-perturbative renormalization program presented in Ref.~\cite{Alexandrou:2017huk}.
It will also have implications in the comparison of the quasi-PDFs and phenomenological estimates for the physical PDFs. 

A possible addition to the present work is the two-loop calculation in dimensional regularization, from which
one can extract the conversion factor between different renormalization schemes, as well as the anomalous 
dimension of the operators.  The conversion factor up to two loops may be applied to non-perturbative data on 
the renormalization functions, to bring them to the $\MSbar$-scheme at a better accuracy. Furthermore, 
knowledge of the two-loop expression for the anomalous dimension in Eq.~(\ref{RRcbar}) will improve the method 
for extracting the linear divergence, and will eliminate systematic uncertainties related to the truncation of the 
conversion factor. 

Another direction that one may follow is the calculation of the one-loop Green's functions at a 
nonzero fermion mass to one loop level. We expect the difference between the finite-mass and massless cases 
to be small given the smooth behavior of the conversion factor, but it would be interesting to investigate, as 
simulations are not exactly at zero renormalized mass. Differences in the renormalization functions of flavor-singlet
and -nonsinglet Wilson line operators will already show up at one loop in the massive case, and at two loops in the
massless case.

Finally, the techniques developed in this work for the renormalization of quasi-PDFs may be inspiring for the
renormalization of Wilson-line fermion operators of different structure, such as staples. Eventually, understanding
the renormalization pattern of such operators may lead to the development of a non-perturbative prescription. 
This will be of high importance for matrix elements of the transverse momentum-dependent parton distributions  
(TMDs) that are currently under investigation for the nucleon and pion in lattice QCD~\cite{Engelhardt:2015xja}. 
Similar to the case of the unpolarized quasi-PDFs, there might be an indication for presence of mixing in certain 
cases~\cite{EG2017}.

\vspace*{1cm}
\centerline{{\bf\Large{{\bf{Acknowledgements}}}}}
\bigskip
We would like to thank the members of ETMC for useful and fruitful discussions. 
In particular, we are thankful to Krzysztof Cichy for discussions in the preparation of this manuscript.
MC acknowledges financial support by the U.S. Department of Energy,
Office of Science, Office of Nuclear Physics, within the framework of
the TMD Topical Collaboration, as well as, by the National Science
Foundation under Grant No. PHY-1714407.

\newpage
\centerline{{\bf\Large{APPENDICES}}}
\appendix
 
\section{Basis of Integrals}
 
The Green's functions for the operators including a Wilson line depend on the
external momentum four-vector, $q_\nu$, and on the length of the Wilson line, $z$\,, in a complicated way which is not
amenable to a closed analytic form. 
In order to present our general results in a compact manner we introduce two classes of integrals:
\begin{itemize}
\item[{\bf{1.}}] $F_1$ - $F_5$: integrals over the Feynman parameter $x$
\item[{\bf{2.}}] $G_1$ - $G_5$: integrals over $x$ and over the parameter $\zeta$ of Eq.~(\ref{Oper})
\end{itemize}
The integrands contain modified Bessel functions of the second kind, $K_0$ and $K_1$. All integrals presented here are
convergent and can be computed numerically. 

\bea
\label{F1}
{F_1(q, z)} &=& {\int_0^{1} dx}\,e^{-i q_\mu x z}\, K_0\left(q\,|z|\, \sqrt{(1-x) x}\right)  \\[1.5ex]
{F_2(q, z)} &=& {\int_0^{1} dx}\, e^{-i q_\mu x z}\,x\,K_0\left(q\,|z|\, \sqrt{(1-x) x}\right) \\[1.5ex]
{F_3(q, z)} &=& {\int_0^{1} dx}\,e^{-i q_\mu x  z}\,(1-x)^2 \,K_0\left(q\,|z|\, \sqrt{(1-x) x}\right) \\[1.5ex]
{F_4(q, z)} &=& {\int_0^{1}  dx}\,e^{-i q_\mu x z}\,\sqrt{(1-x) x}  \,K_1\left(q\,|z|\, \sqrt{(1-x) x}\right) \\[1.5ex]
{F_5(q, z)} &=& {\int_0^{1}  dx}\,e^{-i q_\mu x z}\,(1-x) \sqrt{(1-x) x}  \,K_1\left(q\,|z|\, \sqrt{(1-x) x}\right) 
\eea

\vspace*{0.25cm}
\bea
{G_1(q, z)} &=&   {\int_0^{1} dx \int_0^z d\zeta}\,e^{-i q_\mu x \zeta }\, K_0\left(q\, |\zeta|\,\sqrt{(1-x) x}\right)\\[1.5ex]
{G_2(q, z)} &=&  {\int_0^{1} dx \int_0^z d\zeta}\,e^{-i q_\mu x \zeta }\,x  \,K_0\left(q\, |\zeta|\,\sqrt{(1-x) x}\right)\\[1.5ex]
{G_3(q, z)} &=&  {\int_0^{1} dx\int_0^z d\zeta}\,e^{-i q_\mu x \zeta }\,\zeta\,x\,(1-x)   \,K_0\left(q\, |\zeta|\,\sqrt{(1-x) x}\right) \\[1.5ex]
{G_4(q, z)} &=&  {\int_0^{1} dx \int_0^z d\zeta}\,e^{-i q_\mu x \zeta }\,|\zeta |\,\sqrt{(1-x) x}  \,K_1\left(q\, |\zeta|\,\sqrt{(1-x) x}\right) \\[1.5ex]
{G_5(q, z)} &=&  {\int_0^{1} dx \int_0^z  d\zeta}\,e^{-i q_\mu x \zeta }\,|\zeta |\,x \sqrt{(1-x) x}   \,K_1\left(q\, |\zeta|\,\sqrt{(1-x) x}\right)
\label{G6}
\eea
Here $q \equiv \sqrt{q^2} $. 
Note that the $F$ integrals are dimensionless, $G_1$ - $G_2$ have dimensions of $z$ due to integration over $\zeta$, and 
$G_3$ - $G_5$ have dimensions of $z^2$. The dimensionality of all integrals is combined with the momentum in such a way 
that the conversion factors and Green's functions are dimensionless (see, e.g., Eqs.~(\ref{SP_R})-(\ref{T2Tp2_R})).

\newpage
\section{Renormalized Green's Functions}
 
In this Appendix we present the general expressions for the renormalized Green's functions of the operators given in
Eqs.~(\ref{S}), in the $\MSbar$ scheme. The functions
$ \Lambda^{\rm 1-loop}_\Gamma$ are complex, and have a complicated dependence on the momentum and the length of the Wilson line;
thus, we write them in a compact form, using the list of integrals $F_i(q,z)$ and $G_i(q,z)$ defined in Appendix A.
\bea
\hspace*{-0.5cm}
 \Lambda^{\rm 1-loop}_S = \Lambda^{\rm tree}_S\Bigg[1 + \frac{g^2\,C_f}{16\,\pi^2} \hspace*{-0.4cm}&\Bigg(& \hspace*{-0.4cm}
    8 + 4\gamma_E -\log(16) +8 {F_2}-2 q |z| ({F_4}-2 {F_5})  +  3 \log
    \left(\frac{{\bar{\mu}^2}}{{q^2}}\right)+(\beta +2) \log ({q^2}
    {z^2})\nonumber \\
    &&\hspace*{-0.3cm}+ \beta  \Big[-3 + 2 \gamma_E - \log(4) +2 {F_1}+{q^2} z^2\left(\frac{{F_1-F_2}}{2}-{F_3}\right) +2 \left({q_\mu}^2+{q^2}\right)G_3 - q |z|{F_4}\Big]\nonumber \\
    &&\hspace*{-0.3cm}+ 4 i{q_\mu}
    \Big(z({F_1-F_2}-{F_3}) -{G_1}\Big)\nonumber \\
    &&\hspace*{-0.3cm}+i \beta  {q_\mu} \Big[2 (G_1 -G_2) -q \Big(2 ({G_4}-2{G_5})+ z |z|{F_5}\Big)\Big]\Bigg)\Bigg]
    \label{SP_R}
\eea

\bea
\hspace*{-0.5cm}
 \Lambda^{\rm 1-loop}_{V_\mu} = \Lambda^{\rm tree}_{V_\mu}\Bigg[1 + \frac{g^2\,C_f}{16\,\pi^2} \hspace*{-0.4cm}&\Bigg(& \hspace*{-0.4cm}
8 + 4\gamma_E -\log(16) -4 {F_2}+4 q |z|\,({F_4}-{F_5}) 
+ 3 \log \left(\frac{{\bar{\mu}^2}}{{q^2}}\right)+(\beta +2) \log ({q^2} {z^2}) \nonumber\\
    &&\hspace*{-0.3cm}+\beta  \Big[-3 + 2 \gamma_E - \log(4)+2 {F_1} +2(q^2 + 2{q_\mu}^2) {G_3} -\frac{{q^2} z^2}{2}({F_1-F_2})\Big]\nonumber \\
    &&\hspace*{-0.3cm} +4i {q_\mu} (-2 {G_1}+{G_2}+z({F_3}-{F_1}+{F_2}))\nonumber\\
    &&\hspace*{-0.3cm}  + i\beta  {q_\mu} \Big[4 {G_1}-4 {G_2}+2 z ({F_2}-{F_1}) +2 q (-2 {G_4}+3 {G_5}) \Big]    
    \Bigg)\Bigg] \nonumber \\  [1ex]
&&\hspace*{-3cm}+{\slash \hspace*{-0.2cm}q}\,e^{i {q_\mu} z} \,\frac{g^2\,C_f}{16\,\pi^2} 
    \Bigg[-\frac{4 {q_\mu} |z|}{q}{F_5} +\beta
    {q_\mu} \Big[z^2({F_1-F_2}-{F_3}) -2 {G_3}+\frac{2 |z|}{q}{F_4}\Big] \nonumber\\
    &&\hspace*{-0.5cm}+4i ({G_1}-{G_2}+z({F_3}-{F_1}+{F_2})) \nonumber\\
    &&\hspace*{-0.5cm} +i \beta  \Big[2(G_2{-}{G_1})+2z ({F_1}{-}{F_2})+q (2 {G_4}-2 {G_5}-z |z|{F_5} )\Big]\Bigg]
    \eea

\bea
\hspace*{-0.5cm}
 \Lambda^{\rm 1-loop}_{V_\nu} = \Lambda^{\rm tree}_{V_\nu}\Bigg[1 + \frac{g^2\,C_f}{16\,\pi^2} \hspace*{-0.4cm}&\Bigg(& \hspace*{-0.4cm}
 8 + 4\gamma_E -\log(16) -4 {F_2} +3 \log
    \left(\frac{{\bar{\mu}^2}}{{q^2}}\right)+(\beta +2) \log ({q^2}
    {z^2}) \nonumber\\
 &&\hspace*{-0.3cm} +\beta  \Big[ -3 + 2 \gamma_E - \log(4) +2 {F_1} + 2 ( q^2 + {q_\mu}^2) {G_3} + {q^2} z^2\left( \frac{1}{2}({F_1-F_2})-{F_3}\right)\Big]\nonumber \\ 
  &&\hspace*{-0.3cm} + i\,q_\mu \left(-4  {G_1} +\beta  \Big[2 {G_1}-2
    {G_2}-q (2({G_4}-2{G_5})+z |z|{F_5} )\Big] \right)
    \Bigg)\Bigg] \nonumber \\  [1ex]
&&\hspace*{-3cm}+ {\slash \hspace*{-0.2cm}q}\, q_\nu \,e^{i {q_\mu} z} \,\frac{g^2\,C_f}{16\,\pi^2} 
    \Bigg[\beta \Big[ z^2 ({F_3}-{F_1}+{F_2}) +\frac{2
     |z|}{q}{F_4}\Big]-\frac{4 |z|}{q}{F_5}\Bigg] \nonumber \\ [1ex]
&&\hspace*{-3cm}+\Lambda^{\rm tree}_{V_\mu}\,q_\nu\,\frac{g^2\,C_f}{16\,\pi^2} \Bigg[
    2\beta {q_\mu} {G_3}  + 4i
     (-{G_1}+{G_2}+z({F_3}-{F_1}+{F_2}) )\nonumber\\
    &&\hspace*{-.45cm}+i \beta  \left(2
    ({G_1}-{G_2}+z ({F_2}-{F_1}))+q (-2 {G_4}+2 {G_5}+ z |z|{F_5})\right)   \Bigg]
\qquad (\nu\ne\mu)
\eea

\bea
\hspace*{-1cm}
 \Lambda^{\rm 1-loop}_{{T_{\mu\nu}}} = \Lambda^{\rm tree}_{T_{\mu\nu}}\Bigg[1 + \frac{g^2\,C_f}{16\,\pi^2} \hspace*{-0.4cm}&\Bigg(& \hspace*{-0.4cm}
   8 + 4\gamma_E -\log(16) -8 {F_2}+2 q |z|({F_4}-2 {F_5}) +3 \log
    \left(\frac{{\bar{\mu}^2}}{{q^2}}\right)+(\beta +2) \log ({q^2}
    {z^2}) \nonumber\\
  && \hspace*{-0.4cm}+\beta  \Big[-3 + 2 \gamma_E - \log(4) + 4 q_\mu^2 {G_3} +2 {F_1}
   +{q^2} \left(-\frac{ z^2}{2}({F_1-F_2})+2 {G_3}\right)+ q |z|{F_4}\Big]    \nonumber\\
  &&\hspace*{-0.4cm}  + 4i {q_\mu} \Big(-2
    {G_1}+{G_2}+z({F_3}-{F_1}+{F_2}) \Big)\nonumber\\
 &&\hspace*{-0.4cm} +   2 i\beta  {q_\mu}
    \Big[2 {G_1}-2 {G_2}+z ({F_2}-{F_1})
     +q (-2{G_4}+3 {G_5}) \Big]     \Bigg)\Bigg] \nonumber\\[1ex]
&&\hspace*{-3cm}+ \Big(\Lambda^{\rm tree}_{V_\mu}\, q_\nu - \Lambda^{\rm tree}_{V_\nu}\, q_\mu \Big)\cdot {\slash \hspace*{-0.2cm}q} \, 
\frac{g^2\,C_f}{16\,\pi^2} \,\Bigg[\beta z^2 ({F_1-F_2}-{F_3})  \Bigg] \nonumber \\[1ex]
&&\hspace*{-3cm} +
\Big({\slash \hspace*{-0.2cm}q} \cdot\Lambda^{\rm tree}_{V_\nu}  - 
     \Lambda^{\rm tree}_{V_\nu}\cdot {\slash \hspace*{-0.2cm}q}\Big)
\,\frac{g^2\,C_f}{16\,\pi^2}\Bigg[ 
- \beta  {q_\mu}  {G_3} + 2i ({G_1}-{G_2}) + i\beta \,  (-{G_1}+{G_2}+z ({F_1}-{F_2})) \nonumber\\
&& \hspace*{1.75cm}+ i\beta q ( {G_4}- {G_5}- \frac{1}{2} z |z|{F_5}) \Bigg] 
\qquad\qquad\qquad\qquad (\nu\ne\mu)
\eea

\bea
\hspace*{-1cm}
 \Lambda^{\rm 1-loop}_{T_{\nu\rho}} = 
 \Lambda^{\rm tree}_{T_{\nu\rho}}\Bigg[1 + \frac{g^2\,C_f}{16\,\pi^2} \hspace*{-0.4cm}&\Bigg(& \hspace*{-0.4cm} 
 8 + 4\gamma_E -\log(16) -8 {F_2}+2 q |z| ({F_4}-2  {F_5})  +3 \log
    \left(\frac{{\bar{\mu}^2}}{{q^2}}\right)+(\beta +2) \log ({q^2}
    {z^2}) \nonumber  \\
 &&\hspace*{-0.4cm} +\beta  \Big[-3 + 2 \gamma_E - \log(4) +2 {F_1}+q |z| {F_4}  + 2 \left({q_\mu}^2+{q^2}\right) {G_3}
   +{q^2} z^2\left(\frac{1}{2} ({F_1-F_2}) -{F_3}\right) \Big]\nonumber  \\
 && \hspace*{-0.4cm}     - 4i {q_\mu}
    ({G_1}{+}z({F_1{-}F_2}-{F_3}) )- i \beta  {q_\mu} \Big[{-}2
    {G_1}{+}2 {G_2}{+}q \Big(2{G_4}{-}4{G_5}+{F_5} z
    |z|\Big)\Big]\Bigg)\Bigg]\nonumber\\[1ex]
&&\hspace*{-3cm}
+  \Big(\Lambda^{\rm tree}_{V_\nu}\,q_\rho -\Lambda^{\rm tree}_{V_\rho}\,q_\nu \Big)\cdot {\slash \hspace*{-0.2cm}q} \,\frac{g^2\,C_f}{16\,\pi^2} \,\Bigg[     
-\beta\, z^2  ({F_1-F_2}-{F_3}) \Bigg] \nonumber \\[1ex]
    &&  \hspace*{-3cm}  + \Big(\Lambda^{\rm tree}_{T_{\mu\nu}}\, q_\rho - \Lambda^{\rm tree}_{T_{\mu\rho}}\, q_\nu \Big) \,
     \frac{g^2\,C_f}{16\,\pi^2}  \Bigg[ -2\beta  {q_\mu} {G_3}  + 4i ({G_1}-{G_2}) 
+ 2 i\beta (-{G_1}+{G_2}+z({F_1}-{F_2}))  \nonumber \\
 &&\hspace*{1.65cm} + i\beta q (2 {G_4}-2 {G_5}- z|z|{F_5})\Bigg] 
\qquad\qquad\qquad (\nu\ne\mu,\ \rho\ne\mu)
    \label{T2Tp2_R} 
\eea

\be
\Lambda^{\rm 1-loop}_P = \gamma_5 \,\Lambda^{\rm 1-loop}_S , \qquad\qquad
\Lambda^{\rm 1-loop}_{A_\nu} = \gamma_5 \,\Lambda^{\rm 1-loop}_{V_\nu} \qquad
(\forall \nu)
\ee

\newpage
\bibliographystyle{elsarticle-num}
\bibliography{references}

\begin{thebibliography}{10}
\expandafter\ifx\csname url\endcsname\relax
  \def\url#1{\texttt{#1}}\fi
\expandafter\ifx\csname urlprefix\endcsname\relax\def\urlprefix{URL }\fi
\expandafter\ifx\csname href\endcsname\relax
  \def\href#1#2{#2} \def\path#1{#1}\fi

\bibitem{Ji:2013dva}
X.~Ji, {Parton Physics on a Euclidean Lattice}, Phys. Rev. Lett. 110 (2013)
  262002.
\newblock \href {http://arxiv.org/abs/1305.1539} {\path{arXiv:1305.1539}}.

\bibitem{Lin:2014zya}
H.-W. Lin, J.-W. Chen, S.~D. Cohen, X.~Ji, {Flavor Structure of the Nucleon Sea
  from Lattice QCD}, Phys. Rev. D91 (2015) 054510.
\newblock \href {http://arxiv.org/abs/1402.1462} {\path{arXiv:1402.1462}}.

\bibitem{Alexandrou:2015rja}
C.~Alexandrou, K.~Cichy, V.~Drach, E.~Garcia-Ramos, K.~Hadjiyiannakou,
  K.~Jansen, F.~Steffens, C.~Wiese, {Lattice calculation of parton
  distributions}, Phys. Rev. D92 (2015) 014502.
\newblock \href {http://arxiv.org/abs/1504.07455} {\path{arXiv:1504.07455}}.

\bibitem{Chen:2016utp}
J.-W. Chen, S.~D. Cohen, X.~Ji, H.-W. Lin, J.-H. Zhang, {Nucleon Helicity and
  Transversity Parton Distributions from Lattice QCD}, Nucl. Phys. B911 (2016)
  246--273.
\newblock \href {http://arxiv.org/abs/1603.06664} {\path{arXiv:1603.06664}}.

\bibitem{Alexandrou:2016jqi}
C.~Alexandrou, K.~Cichy, M.~Constantinou, K.~Hadjiyiannakou, K.~Jansen,
  F.~Steffens, C.~Wiese, {Updated Lattice Results for Parton Distributions},
  Phys. Rev. D96~(1) (2017) 014513.
\newblock \href {http://arxiv.org/abs/1610.03689} {\path{arXiv:1610.03689}}.

\bibitem{PhysRev.175.1580}
S.~Mandelstam, Feynman rules for electromagnetic and yang-mills fields from the
  gauge-independent field-theoretic formalism, Phys. Rev. 175 (1968)
  1580--1603.

\bibitem{POLYAKOV1979247}
A.~Polyakov, String representations and hidden symmetries for gauge fields,
  Physics Letters B 82~(2) (1979) 247 -- 250.

\bibitem{MAKEENKO1979135}
Y.~Makeenko, A.~Migdal, Exact equation for the loop average in multicolor qcd,
  Physics Letters B 88~(1) (1979) 135 -- 137.

\bibitem{Witten1989629}
E.~Witten, Gauge theories and integrable lattice models, Nuclear Physics B
  322~(3) (1989) 629 -- 697.

\bibitem{Dotsenko:1979wR}
V.~S. Dotsenko, S.~N. Vergeles, {Renormalizability of Phase Factors in the
  Nonabelian Gauge Theory}, Nucl. Phys. B169 (1980) 527--546.

\bibitem{Brandt:1981kf}
R.~A. Brandt, F.~Neri, M.-a. Sato, {Renormalization of Loop Functions for All
  Loops}, Phys. Rev. D24 (1981) 879.

\bibitem{DiGiacomo:1992hhp}
A.~Di~Giacomo, H.~Panagopoulos, {Field strength correlations in the QCD
  vacuum}, Phys. Lett. B285 (1992) 133--136.

\bibitem{DiGiacomo:1996bbx}
A.~Di~Giacomo, E.~Meggiolaro, H.~Panagopoulos, {Gauge invariant field
  correlators in QCD at finite temperature}, Nucl. Phys. B483 (1997) 371--382.
\newblock \href {http://arxiv.org/abs/hep-lat/9603018}
  {\path{arXiv:hep-lat/9603018}}.

\bibitem{DiGiacomo:2000irz}
A.~Di~Giacomo, H.~G. Dosch, V.~I. Shevchenko, {\relax Yu}.~A. Simonov, {Field
  correlators in QCD: Theory and applications}, Phys. Rept. 372 (2002)
  319--368.
\newblock \href {http://arxiv.org/abs/hep-ph/0007223}
  {\path{arXiv:hep-ph/0007223}}.

\bibitem{Simonov:2007dn}
{\relax Yu}.~A. Simonov, {Nonperturbative equation of state of quark-gluon
  plasma}, Annals Phys. 323 (2008) 783.
\newblock \href {http://arxiv.org/abs/hep-ph/0702266}
  {\path{arXiv:hep-ph/0702266}}.

\bibitem{Giordano:2009vs}
M.~Giordano, E.~Meggiolaro, {Instanton effects on Wilson-loop correlators: a
  new comparison with numerical results from the lattice}, Phys. Rev. D81
  (2010) 074022.
\newblock \href {http://arxiv.org/abs/0910.4505} {\path{arXiv:0910.4505}}.

\bibitem{Xiong:2013bka}
X.~Xiong, X.~Ji, J.-H. Zhang, Y.~Zhao, {One-loop matching for parton
  distributions: Nonsinglet case}, Phys. Rev. D90~(1) (2014) 014051.
\newblock \href {http://arxiv.org/abs/1310.7471} {\path{arXiv:1310.7471}}.

\bibitem{Ma:2014jla}
Y.-Q. Ma, J.-W. Qiu, {Extracting Parton Distribution Functions from Lattice QCD
  Calculations, }\href {http://arxiv.org/abs/1404.6860}
  {\path{arXiv:1404.6860}}.

\bibitem{Ma:2014jga}
Y.-Q. Ma, J.-W. Qiu, {QCD Factorization and PDFs from Lattice QCD Calculation},
  Int. J. Mod. Phys. Conf. Ser. 37 (2015) 1560041.
\newblock \href {http://arxiv.org/abs/1412.2688} {\path{arXiv:1412.2688}}.

\bibitem{Chen:2016fxx}
J.-W. Chen, X.~Ji, J.-H. Zhang, {Improved quasi parton distribution through
  Wilson line renormalization}, Nucl. Phys. B915 (2017) 1--9.
\newblock \href {http://arxiv.org/abs/1609.08102} {\path{arXiv:1609.08102}}.

\bibitem{Li:2016amo}
H.-n. Li, {Nondipolar Wilson links for quasiparton distribution functions},
  Phys. Rev. D94~(7) (2016) 074036.
\newblock \href {http://arxiv.org/abs/1602.07575} {\path{arXiv:1602.07575}}.

\bibitem{Musch:2011er}
B.~U. Musch, P.~Hagler, M.~Engelhardt, J.~W. Negele, A.~Schafer, {Sivers and
  Boer-Mulders observables from lattice QCD}, Phys. Rev. D85 (2012) 094510.
\newblock \href {http://arxiv.org/abs/1111.4249} {\path{arXiv:1111.4249}}.

\bibitem{Engelhardt:2015xja}
M.~Engelhardt, P.~Hagler, B.~Musch, J.~Negele, A.~Schafer, {Lattice QCD study
  of the Boer-Mulders effect in a pion}, Phys. Rev. D93~(5) (2016) 054501.
\newblock \href {http://arxiv.org/abs/1506.07826} {\path{arXiv:1506.07826}}.

\bibitem{Radyushkin:2016hsy}
A.~Radyushkin, {Nonperturbative Evolution of Parton Quasi-Distributions}, Phys.
  Lett. B767 (2017) 314--320.
\newblock \href {http://arxiv.org/abs/1612.05170} {\path{arXiv:1612.05170}}.

\bibitem{Radyushkin:2017ffo}
A.~Radyushkin, {Target Mass Effects in Parton Quasi-Distributions}, Phys. Lett.
  B770 (2017) 514--522.
\newblock \href {http://arxiv.org/abs/1702.01726} {\path{arXiv:1702.01726}}.

\bibitem{Musch:2010ka}
B.~U. Musch, P.~Hagler, J.~W. Negele, A.~Schafer, {Exploring quark transverse
  momentum distributions with lattice QCD}, Phys. Rev. D83 (2011) 094507.
\newblock \href {http://arxiv.org/abs/1011.1213} {\path{arXiv:1011.1213}}.

\bibitem{Ishikawa:2016znu}
T.~Ishikawa, Y.-Q. Ma, J.-W. Qiu, S.~Yoshida, {Practical quasi parton
  distribution functions, }\href {http://arxiv.org/abs/1609.02018}
  {\path{arXiv:1609.02018}}.

\bibitem{Monahan:2016bvm}
C.~Monahan, K.~Orginos, {Quasi parton distributions and the gradient flow},
  JHEP 03 (2017) 116.
\newblock \href {http://arxiv.org/abs/1612.01584} {\path{arXiv:1612.01584}}.

\bibitem{Carlson:2017gpk}
C.~E. Carlson, M.~Freid, {Lattice corrections to the quark quasidistribution at
  one-loop}, Phys. Rev. D95~(9) (2017) 094504.
\newblock \href {http://arxiv.org/abs/1702.05775} {\path{arXiv:1702.05775}}.

\bibitem{Briceno:2017cpo}
R.~A. Briceno, M.~T. Hansen, C.~J. Monahan, {Role of the Euclidean signature in
  lattice calculations of quasidistributions and other nonlocal matrix
  elements}, Phys. Rev. D96~(1) (2017) 014502.
\newblock \href {http://arxiv.org/abs/1703.06072} {\path{arXiv:1703.06072}}.

\bibitem{Xiong:2017jtn}
X.~Xiong, T.~Luu, U.-G. Meissner, {Quasi-Parton Distribution Function in
  Lattice Perturbation Theory, }\href {http://arxiv.org/abs/1705.00246}
  {\path{arXiv:1705.00246}}.

\bibitem{Sheikholeslami:1985ij}
B.~Sheikholeslami, R.~Wohlert, {Improved Continuum Limit Lattice Action for QCD
  with Wilson Fermions}, Nucl. Phys. B259 (1985) 572.

\bibitem{Horsley:2004mx}
R.~Horsley, H.~Perlt, P.~E.~L. Rakow, G.~Schierholz, A.~Schiller, {One-loop
  renormalisation of quark bilinears for overlap fermions with improved gauge
  actions}, Nucl. Phys. B693 (2004) 3--35, [Erratum: Nucl.
  Phys.B713,601(2005)].
\newblock \href {http://arxiv.org/abs/hep-lat/0404007}
  {\path{arXiv:hep-lat/0404007}}.

\bibitem{Constantinou:2015ela}
M.~Constantinou, M.~Costa, R.~Frezzotti, V.~Lubicz, G.~Martinelli, D.~Meloni,
  H.~Panagopoulos, S.~Simula, {Renormalization of the chromomagnetic operator
  on the lattice}, Phys. Rev. D92~(3) (2015) 034505.
\newblock \href {http://arxiv.org/abs/1506.00361} {\path{arXiv:1506.00361}}.

\bibitem{Alexandrou:2016ekb}
C.~Alexandrou, M.~Constantinou, K.~Hadjiyiannakou, K.~Jansen, H.~Panagopoulos,
  C.~Wiese, {Gluon momentum fraction of the nucleon from lattice QCD}, Phys.
  Rev. D96~(5) (2017) 054503.
\newblock \href {http://arxiv.org/abs/1611.06901} {\path{arXiv:1611.06901}}.

\bibitem{Alexandrou:2017huk}
C.~Alexandrou, K.~Cichy, M.~Constantinou, K.~Hadjiyiannakou, K.~Jansen,
  H.~Panagopoulos, F.~Steffens, {A complete non-perturbative renormalization
  prescription for quasi-PDFs}, Nucl. Phys. B923 (2017) 394--415.
\newblock \href {http://arxiv.org/abs/1706.00265} {\path{arXiv:1706.00265}}.

\bibitem{Constantinou:2009tr}
M.~Constantinou, V.~Lubicz, H.~Panagopoulos, F.~Stylianou, {O(a**2) corrections
  to the one-loop propagator and bilinears of clover fermions with Symanzik
  improved gluons}, JHEP 10 (2009) 064.
\newblock \href {http://arxiv.org/abs/0907.0381} {\path{arXiv:0907.0381}}.

\bibitem{Chetyrkin:1999pq}
K.~G. Chetyrkin, A.~Retey, {Renormalization and running of quark mass and field
  in the regularization invariant and MS-bar schemes at three loops and four
  loops}, Nucl. Phys. B583 (2000) 3--34.
\newblock \href {http://arxiv.org/abs/hep-ph/9910332}
  {\path{arXiv:hep-ph/9910332}}.

\bibitem{tHooft:1973wag}
G.~'t~Hooft, M.~J.~G. Veltman, {DIAGRAMMAR}, NATO Sci. Ser. B 4 (1974)
  177--322.

\bibitem{Gracey:2003yr}
J.~A. Gracey, {Three loop anomalous dimension of nonsinglet quark currents in
  the RI-prime scheme}, Nucl. Phys. B662 (2003) 247--278.
\newblock \href {http://arxiv.org/abs/hep-ph/0304113}
  {\path{arXiv:hep-ph/0304113}}.

\bibitem{Dorn:1986dt}
H.~Dorn, {Renormalization of Path Ordered Phase Factors and Related Hadron
  Operators in Gauge Field Theories}, Fortsch. Phys. 34 (1986) 11--56.

\bibitem{Chetyrkin:2003vi}
K.~G. Chetyrkin, A.~G. Grozin, {Three loop anomalous dimension of the heavy
  light quark current in HQET}, Nucl. Phys. B666 (2003) 289--302.
\newblock \href {http://arxiv.org/abs/hep-ph/0303113}
  {\path{arXiv:hep-ph/0303113}}.

\bibitem{Buras:1989xd}
A.~J. Buras, P.~H. Weisz, {QCD Nonleading Corrections to Weak Decays in
  Dimensional Regularization and 't Hooft-Veltman Schemes}, Nucl. Phys. B333
  (1990) 66--99.

\bibitem{Patel:1992vu}
A.~Patel, S.~R. Sharpe, {Perturbative corrections for staggered fermion
  bilinears}, Nucl. Phys. B395 (1993) 701--732.
\newblock \href {http://arxiv.org/abs/hep-lat/9210039}
  {\path{arXiv:hep-lat/9210039}}.

\bibitem{Larin:1993tp}
S.~A. Larin, J.~A.~M. Vermaseren, {The Three loop QCD Beta function and
  anomalous dimensions}, Phys. Lett. B303 (1993) 334--336.
\newblock \href {http://arxiv.org/abs/hep-ph/9302208}
  {\path{arXiv:hep-ph/9302208}}.

\bibitem{Larin:1993tq}
S.~A. Larin, {The Renormalization of the axial anomaly in dimensional
  regularization}, Phys. Lett. B303 (1993 [See also hep-ph/9302240 for an
  additional Section]) 113--118.
\newblock \href {http://arxiv.org/abs/hep-ph/9302240}
  {\path{arXiv:hep-ph/9302240}}.

\bibitem{Skouroupathis:2008mf}
A.~Skouroupathis, H.~Panagopoulos, {Two-loop renormalization of vector,
  axial-vector and tensor fermion bilinears on the lattice}, Phys. Rev. D79
  (2009) 094508.
\newblock \href {http://arxiv.org/abs/0811.4264} {\path{arXiv:0811.4264}}.

\bibitem{Constantinou:2013pba}
M.~Constantinou, M.~Costa, H.~Panagopoulos, {Perturbative renormalization
  functions of local operators for staggered fermions with stout improvement},
  Phys. Rev. D88 (2013) 034504.
\newblock \href {http://arxiv.org/abs/1305.1870} {\path{arXiv:1305.1870}}.

\bibitem{Gracey:2000am}
J.~A. Gracey, {Three loop MS-bar tensor current anomalous dimension in QCD},
  Phys. Lett. B488 (2000) 175--181.
\newblock \href {http://arxiv.org/abs/hep-ph/0007171}
  {\path{arXiv:hep-ph/0007171}}.

\bibitem{Kawai:1980ja}
H.~Kawai, R.~Nakayama, K.~Seo, {Comparison of the Lattice Lambda Parameter with
  the Continuum Lambda Parameter in Massless QCD}, Nucl. Phys. B189 (1981)
  40--62.

\bibitem{Sommer:2015hea}
R.~Sommer, {Non-perturbative Heavy Quark Effective Theory: Introduction and
  Status}, Nucl. Part. Phys. Proc. 261-262 (2015) 338--367.
\newblock \href {http://arxiv.org/abs/1501.03060} {\path{arXiv:1501.03060}}.

\bibitem{Alexandrou:2012mt}
C.~Alexandrou, M.~Constantinou, T.~Korzec, H.~Panagopoulos, F.~Stylianou,
  {Renormalization constants of local operators for Wilson type improved
  fermions}, Phys. Rev. D86 (2012) 014505.
\newblock \href {http://arxiv.org/abs/1201.5025} {\path{arXiv:1201.5025}}.

\bibitem{Jaffe:1991ra}
R.~L. Jaffe, X.-D. Ji, {Chiral odd parton distributions and Drell-Yan
  processes}, Nucl. Phys. B375 (1992) 527--560.

\bibitem{Gockeler:2010yr}
M.~Gockeler, et~al., {Perturbative and Nonperturbative Renormalization in
  Lattice QCD}, Phys. Rev. D82 (2010) 114511, [Erratum: Phys.
  Rev.D86,099903(2012)].
\newblock \href {http://arxiv.org/abs/1003.5756} {\path{arXiv:1003.5756}}.

\bibitem{Constantinou:2013ada}
M.~Constantinou, M.~Costa, M.~Gockeler, R.~Horsley, H.~Panagopoulos, H.~Perlt,
  P.~E.~L. Rakow, G.~Schierholz, A.~Schiller, {Perturbatively improving
  regularization-invariant momentum scheme renormalization constants}, Phys.
  Rev. D87~(9) (2013) 096019.
\newblock \href {http://arxiv.org/abs/1303.6776} {\path{arXiv:1303.6776}}.

\bibitem{Constantinou:2014fka}
M.~Constantinou, R.~Horsley, H.~Panagopoulos, H.~Perlt, P.~E.~L. Rakow,
  G.~Schierholz, A.~Schiller, J.~M. Zanotti, {Renormalization of local
  quark-bilinear operators for $N_f$=3 flavors of stout link nonperturbative
  clover fermions}, Phys. Rev. D91~(1) (2015) 014502.
\newblock \href {http://arxiv.org/abs/1408.6047} {\path{arXiv:1408.6047}}.

\bibitem{Alexandrou:2015sea}
C.~Alexandrou, M.~Constantinou, H.~Panagopoulos, {Renormalization functions for
  Nf=2 and Nf=4 twisted mass fermions}, Phys. Rev. D95~(3) (2017) 034505.
\newblock \href {http://arxiv.org/abs/1509.00213} {\path{arXiv:1509.00213}}.

\bibitem{EG2017}
M.~Engelhardt, R.~Gupta, private communication.

\end{thebibliography}

\end{document}